\documentclass[twocolumn,floatfix,showpacs,amsmath,amssymb,pra]{revtex4}
\usepackage{times}
\usepackage{graphicx}
\usepackage{dcolumn}
\usepackage{bm}

\begin{document}

\title{Quantum versus classical statistical dynamics of an ultracold Bose gas}
\author{J\"urgen Berges}
\thanks{email:Juergen.Berges@physik.tu-darmstadt.de}
\affiliation{Institut f\"ur Kernphysik, Technische Universit\"at
Darmstadt, Schlossgartenstrasse 9, 64289 Darmstadt, Germany}
\author{Thomas Gasenzer}
\thanks{email:T.Gasenzer@thphys.uni-heidelberg.de}
\affiliation{Institut f\"ur Theoretische Physik, Universit\"at
Heidelberg, Philosophenweg 16, 69120 Heidelberg, Germany}

\begin{abstract}
\noindent We investigate the conditions under which quantum
fluctuations are relevant for the quantitative interpretation of
experiments with ultracold Bose gases. This requires to go beyond
the description in terms of the Gross-Pitaevskii and
Hartree-Fock-Bogoliubov mean-field theories, which can be obtained
as classical (statistical) field-theory approximations of the
quantum many-body problem. We employ functional-integral
techniques based on the two-particle irreducible (2PI) effective
action. The role of quantum fluctuations is studied within the
nonperturbative 2PI $1/\cal N$ expansion to next-to-leading order.
At this accuracy level memory-integrals enter the dynamic
equations, which differ for quantum and classical statistical
descriptions. This can be used to obtain a 'classicality'
condition for the many-body dynamics. We exemplify this condition
by studying the nonequilibrium evolution of a 1D Bose gas of
sodium atoms, and discuss some distinctive properties of quantum
versus classical statistical dynamics.
\end{abstract}
\pacs{03.75.Kk, 03.75.Nt, 05.30.-d, 05.70.Ln, 11.15.Pg
\hfill HD--THEP--07--05}

\maketitle
\section{Introduction}
\label{sec:intro}

The preparation of ultracold atomic Bose and Fermi gases in
various trapping environments allows to study in a precise way
important aspects of quantum many-body
dynamics~\cite{Cornell2002a,Leggett2001a,Dalfovo1999a,Ketterle1999a,Ketterle1999b,Regal2004b,Stoof1999b,Grimm2007a}.
For this reason the field has attracted in recent years
researchers from a variety of physical disciplines, ranging from
condensed-matter to high-energy particle physics and even
cosmology. In past experiments with Bose-Einstein condensates of
dilute gases it has been found that these are, in many cases,
approximately described by a complex scalar field which solves the
Gross-Pitaevskii equation (GPE)~\cite{Gross1961a,Pitaevskii1961a}.
Despite the fact that the first-order coherence reflected by this
equation has its origin in the quantum nature of the Bose
condensation phenomenon, the GPE arises as the classical
field-theory approximation of the underlying quantum many-body
problem. It thus neglects all quantum statistical fluctuations
contributing to the dynamics of the scalar field. However, it is
the role of these quantum statistical fluctuations which is of
central importance for our quantitative understanding of a wealth
of phenomena described by quantum many-body dynamics. Accordingly,
experiments which are sensitive to fluctuations are crucial to
test our theoretical understanding of complex many-body problems.

If fluctuations are relevant then the quantitative interpretation
of the data typically requires nonperturbative theoretical
descriptions, which often have to take into account nonequilibrium
dynamics to match realistic experimental situations. Two cases
should be distinguished in this context: If the real-time dynamics
of a Bose gas is dominated by classical statistical fluctuations
then it can be well approximated by a large number of numerical
integrations of the classical field equation (GPE) and Monte Carlo
sampling techniques~\cite{Davis2002a,Davis2002b,Kohl2002a}. This
takes into account nonperturbative dynamics, however, neglects all
quantum corrections. A corresponding classical statistical
description does not exist for fermions. The other case concerns
dynamics where quantum fluctuations are relevant. It is an
important challenge to quantitatively determine the role of
quantum fluctuations, thus predicting signatures for the detection
of genuine quantum effects in an experiment. Here, we address this
question for a Bose gas.

In fact, many experiments concerning ultracold Bose gases fall
short of being sensitive to quantum statistical fluctuations, and
can be accurately described by the GPE. The importance of
classical statistical fluctuations can rise if the gas is
sufficiently dense. A combination of low densities and strong
self-interaction can lead to enhanced quantum fluctuations as
compared to classical statistical fluctuations. Zero-energy
scattering resonances, particularly the so called magnetic
Feshbach resonances
\cite{Stwalley1976b,Tiesinga1992a,Tiesinga1993a,Burnett1998a,Kohler2006a}
so far have played a leading role in the creation of strong
interactions in degenerate atomic quantum gases. Near a Feshbach
resonance, the scattering of, e.g., a pair of Bose-condensed
atoms, whose relative energy is very close to zero, can be
described by a strongly enhanced $s$-wave scattering length $a$.
Present-day experimental techniques allow for resonance-enhanced
scattering lengths larger than the mean interatomic distance
$(N/V)^{-1/3}$ in the gas. As a consequence, the diluteness
parameter $a^3 N/V$ is larger than one. The Bose-Einstein
condensate is no longer in the collisionless regime, it represents
a strongly interacting system. Feshbach resonances have gained a
strong practical importance for fermionic gases, where losses are
suppressed in the unitary limit \cite{Petrov2005a} and where they
allow to study the transition from a phase of Bose-condensed
molecules to a BCS type superfluid
\cite{Regal2004b,Zwierlein2004a,Bartenstein2004a,Grimm2007a}.

One- and two-dimensional traps
\cite{Schmiedmayer2000a,Pitaevskii2003a} as well as optical
lattices \cite{Jaksch1998a,Bloch2004a} allow to realize strongly
correlated many-body states of atoms. In an optical lattice,
strong effective interactions can be induced by suppressing the
hopping between adjacent lattice sites and thus increasing the
weight of the interaction relative to the kinetic energy
\cite{Jaksch1998a,vanOosten2001a}. This leads, in the limit of
near-zero hopping or strong interactions, to a Mott-insulating
state \cite{Greiner2002a}. In a one-dimensional trap, the gas
enters the so-called Tonks-Girardeau regime, if the dimensionless
interaction parameter $\gamma=g_\mathrm{1D}m V/(\hbar^2 N)$ is
much larger than one
\cite{Tonks1936a,Girardeau1960a,Paredes2004a}. Here,
$g_\mathrm{1D}$ is the coupling parameter of the one-dimensional
gas, e.g., $g_\mathrm{1D}=2\hbar^2a/(ml_\perp^2)$ for a
cylindrical trap with transverse harmonic oscillator length
$l_\perp$. In the Tonks-Girardeau limit $\gamma\to\infty$ the
atoms can no longer pass each other and behave in many respects
like a one-dimensional ideal Fermi gas \cite{Pitaevskii2003a}.

The theory of the full nonperturbative real-time quantum dynamics
is in general a demanding problem. Already the description of
weakly correlated many-body dynamics suffers from the problem that
it requires summations of infinite series of perturbative
processes. These summations can be efficiently taken into account
using functional-integral techniques for the quantum field theory,
which are based on the two-particle irreducible (2PI) effective
action \cite{Luttinger1960a,Baym1962a,Cornwall1974a}. Much
progress has recently been achieved using nonperturbative
expansions of the 2PI effective action to next-to-leading order in
the number of field components \cite{Berges2002a,Aarts2002b}. This
has been employed in the context of ultracold quantum gases in
Ref.~\cite{Rey2004a,Gasenzer2005a,Temme2006a}. It has previously
been successfully used to study far-from-equilibrium dynamics and
thermalization in relativistic bosonic
\cite{Berges2002a,Berges2003b,Cooper2003a,Arrizabalaga2004a} and
fermionic \cite{Berges2003a,Berges2004b} theories. For an
introductory review see Ref.~\cite{Berges2005a}. In
Ref.~\cite{Aarts2002a} the approach has been used to compare
quantum and classical statistical nonequilibrium dynamics for a
relativistic scalar field theory in the absence of a field
expectation value. Since the dynamic equations for the quantum and
classical correlation functions differ only by few characteristic
terms, this can be used to derive a 'classicality' condition for
many-body dynamics \cite{Aarts2002a,Berges:2002wf}.

Here we extend the analytic discussion including a non-zero
macroscopic field and apply it to a non-relativistic theory for an
ultracold gas with, in the quantum case, bosonic statistics. For
our comparison of quantum and classical dynamics we employ the
functional-integral approach of the 2PI effective action. We
recall that the difference between the quantum and classical
statistical theory can be expressed in terms of interaction vertex
terms for the quantum theory which are absent in the classical
statistical theory. As a consequence, the classical generating
functional is characterized by an important reparametrization
property, which allows one to scale out the dependence of the
dynamics on the scattering length $a$. As a consequence, for the
classical dynamics the effects of a larger self-interaction can
always be compensated by a smaller density. It is shown that
quantum corrections violate this invariance property. They become
of increasing importance with growing scattering length or reduced
density. This is used to derive a condition which may be used for
experimenters to find signatures of quantum fluctuations when
preparing and probing the dynamics of ultracold gases. This
condition is not based on thermal equilibrium assumptions and
holds also for far-from-equilibrium dynamics. As an application,
we study the thermalization of a homogeneous one-dimensional
ultracold Bose gas starting from a far-from-equilibrium initial
state following Ref.~\cite{Gasenzer2005a}. We compare quantum and
classical evolution and demonstrate the validity of our criterion
for the nonequilibrium quantum field theory to be well
approximated by its classical counterpart. To round off the
analysis we discuss some distinctive properties of quantum versus
classical statistical dynamics. For instance, the decay of
correlations with the initial state happens faster in the
classical statistical theory for the one-dimensional Bose gas.
Including quantum corrections the system remembers longer the
details about the initial conditions.

The use of functional methods to describe the dynamics of
classical correlations dates back to the work of Hopf in the
context of statistical hydrodynamics \cite{Hopf1952a}. A field
theory for the description of classical fluctuations in terms of
noncommutative classical fields was first suggested by Martin,
Siggia, and Rose (MSR) \cite{Martin1973a} and has been extensively
used in critical dynamics near equilibrium \cite{Hohenberg1977a}.
This theory has been reformulated later in terms of Lagrangian
field theory employing functional methods
\cite{Phythian1975a,DeDominicis1976a,Janssen1976a,Bausch1976a,Phythian1977a,DeDominicis1978a}.
In these field theoretical approaches to classical statistics, a
doubling of the degrees of freedom occurs. For example, in the
generating functional for Green's functions, besides each field
appearing in the fundamental Lagrangian, a second `response' field
is integrated over. The functional integral approach to quantum
field dynamics developed by Schwinger and Keldysh employs a closed
time path (CTP) contour \cite{Schwinger1961a,Keldysh1964a} in the
time-ordered exponential integral. The doubling of fields in the
MSR and Lagrangian approaches to classical dynamics corresponds to
the fields evaluated separately on the two branches of the
Schwinger-Keldysh CTP~\cite{Chou1980a,Chou1985a,Blagoev2001a}.
Implications of the differences between the classical and quantum
vertices, similar to the case considered in this article, have
been discussed, for other theories, e.g.\ in
Refs.~\cite{Aarts1997a,Buchmuller1997a,Wetterich:1997rp,Aarts1998a,Cooper2001a,Blagoev2001a,Jeon2005a}.

Our article is organized as follows: In Section \ref{sec:ClvsQ} we
recall the functional description of quantum as well as classical
statistical non-equilibrium dynamics and use this to construct the
respective 2PI effective actions.  We then derive the time
evolution equations and compare the nonperturbative expansion in
the numbers of field components to next-to-leading order for the
non-relativistic quantum and classical statistical theory. In
Section \ref{sec:NumRes} we present numerical results for the
quantum and classical evolutions of an ultracold one-dimensional
Bose gas. Our conclusions are drawn in Section \ref{sec:Concl}. In
an appendix we provide some details of employed initial-state density matrices.

\section{Classical versus quantum dynamics of an interacting gas}
\label{sec:ClvsQ}

We consider an ultracold gas of atoms with bosonic statistics. At
sufficiently large phase-space densities the system can undergo a
phase transition and form a Bose-Einstein condensate, given that
dimensionality and trapping geometry fulfill the necessary
conditions. For a dilute gas, i.e., if the atomic distance is much
smaller than the characteristic length scale of the interactions,
typically the $s$-wave scattering length $a$, the system may be
described by a complex scalar field theory.

We consider such a non-relativistic quantum field theory for a
complex-valued field $\varphi(x)$. This fluctuating field is
characterized by the Lagrangian density
\begin{align}
  {\cal L}(x)
  &= \frac{i}{2}\left[\varphi^*(x)\partial_{x_0}\varphi(x)
                     -\varphi(x)\partial_{x_0}\varphi^*(x)\right]
  \nonumber\\
  &\
     -\frac{1}{2m}\partial_{i}\varphi^*(x)\partial_{i}\varphi(x)
     -V(x)\varphi^*(x)\varphi(x)
  \nonumber\\
  &\
     -\frac{g}{2}(\varphi^*(x)\varphi(x))^2
\label{eq:ClassLD}
\end{align}
in the defining functional integral for correlation functions as
described below. We use units where $\hbar=1$. The space-time
variable is $x=(x_0=t,{\bf x})$. It is summed over double indices
$i=1,...,d$ for $d$ spatial dimensions and
$\partial_i\equiv\partial/\partial x_i$. Here $V$ denotes an
external potential, and $g$ a real-valued coupling constant. The
Euler-Lagrange equation of motion derived from (\ref{eq:ClassLD})
reads
\begin{align}
  i\partial_{x_0}\varphi(x)
  &=
  \left[-\frac{\partial_{i}^2}{2m}+V(x)+g|\varphi(x)|^2\right]\varphi(x).
\label{eq:eomClassComplphi}
\end{align}
In the context of the physics of quantum gases of
indistinguishable Bosons, Eq.
(\ref{eq:eomClassComplphi}) is the well known Gross-Pitaevskii
equation if the fluctuating field $\varphi(x)$ is identified with
its quantum statistical average $\langle\varphi(x)\rangle$
\cite{Gross1961a,Pitaevskii1961a}. This equation approximately
describes the time evolution of an inherently quantum system, a
Bose-Einstein condensed ultracold gas, neglecting all quantum
statistical fluctuations contributing to the dynamics of the
scalar field. As such, the Gross-Pitaevskii equation is a
classical field equation and the inclusion of quantum and
statistical fluctuations beyond the classical field approximation
is described in the following.

For this we recall in this section the field theoretical
formulation of the many-body quantum and classical statistical
time evolution. This gives that the dynamic equations for
correlation functions differ only by certain terms in the quantum
equations which are absent in the classical ones. As a side
result, one recovers that the dynamics in the
Hartree-Fock-Bogoliubov (HFB)
\cite{Hartree1928a,Fock1930a,Bogoliubov1947a} approximation is the
same for quantum and classical statistical descriptions for same
initial conditions. Moreover, the differences between the quantum
and classical dynamical equations are identified in the
non-perturbative 2PI $1/\cal N$ approximation which goes far
beyond HFB.

While the field in Eq.~(\ref{eq:ClassLD}) is assumed to be
complex-valued, we will, in the following, switch to a
representation of the field in terms of its real and imaginary
part, $\varphi=(\varphi_1+i\varphi_2)/\sqrt{2}$, where the
classical action reads \cite{Gasenzer2005a}
\begin{eqnarray}
  S[\varphi]
  &=& \frac{1}{2}\int_{x y} \varphi_a(x) iG_{0,ab}^{-1}(x,y)\varphi_b(y)
  \nonumber\\
  &&  -\frac{g}{8} \int_{x} \varphi_a(x)\varphi_a(x)\varphi_b(x)\varphi_b(x) , \label{eq:Sq}
\end{eqnarray}
with $\int_x \equiv \int {\mathrm d} x_0 \int {\mathrm d}^d x$.
The free classical inverse propagator is given by
\begin{eqnarray}
\label{eq:invfreeProp}
  iG^{-1}_{0,ab}(x,y)
  &=& \delta(x-y)\left[-i\sigma^2_{ab}\partial_{x_0}
   -H_\mathrm{1B}(x)\delta_{ab}
     \right],
\end{eqnarray}
where $H_\mathrm{1B}(x)=-\partial^2_i/2m+V(x)$ denotes the
single-particle Hamiltonian with interaction potential $V(x)$ and
\begin{align}
\label{eq:PauliMat2}
  \sigma^2 = \left(\begin{array}{cc}0&-i\\i&0\end{array}\right)
\end{align}
the Pauli matrix in field-index space. Summation over double
indices $a,b,c \ldots= 1,2$ is implied. Many of the following
formal derivations are independent of the detailed form
(\ref{eq:invfreeProp}) of $G_0^{-1}$ and, in particular, equally
valid for relativistic field theories. Note however that as
$G_0^{-1}$ in Eq.~(\ref{eq:invfreeProp}) contains only a
first-order time derivative, the canonically conjugate field, in
the non-relativistic theory, is
\begin{align}
  \pi_a(x)
  =\frac{\delta S[\varphi]}{\delta(\partial_{x_0}\varphi_a(x))}
  = i\sigma^2_{ab}\varphi_b(x) ,
  \label{eq:canmom}
\end{align}
in contrast to the relativistic case where the canonical momentum
equals the time derivative of the field.

\subsection{Quantum statistical dynamics}
\label{sec:FIntQ}
%
\subsubsection{Generating functional}
For a given initial-state density matrix $\rho_D(t_0)$, which may
characterize a system also far from equilibrium, all information
about the quantum field theory is contained in the generating
functional for correlation functions:
\begin{eqnarray}
\label{eq:definingZneq}
  Z[J,K;\rho_D]
  &=& {\rm Tr}\Big[\rho_D(t_0)\, {\cal T}_{\cal C}
  \exp\Big\{i \Big(\int_{x,{\cal C}}\! J_a^{\cal C}(x) \Phi_a(x)
  \nonumber\\
  && +\ \frac{1}{2} \int_{x y,{\cal C}}\!
  \Phi_a(x)K_{ab}^{\cal C}(x,y)\Phi_b(y)\Big)\Big\}\Big]  \, ,
\end{eqnarray}
with Heisenberg field operators $\Phi_a(x)$ which obey for the
non-relativistic theory the commutation relations
\begin{align}
  [{\Phi}_a(t,{\bf x}),{\Phi}_b(t,{\bf y})]=-\sigma^2_{ab}
  \delta({\bf x} -{\bf y}).
  \label{eq:BoseCommutator}
\end{align}
In Eq.~(\ref{eq:definingZneq}), ${\cal T}_{\cal C}$ denotes
time-ordering along the closed time path $\cal C$ leading from
the initial time $t_0$ along the real time axis to some arbitrary
time $t$ and back to $t_0$, with $\int_{x,{\cal C}} \equiv
\int_{\cal C} {\mathrm d} x_0 \int {\mathrm d}^d x$. Contour time
ordering along this path corresponds to usual time ordering along
the forward piece ${\cal C}^+$ and antitemporal ordering on the
backward piece ${\cal C}^-$. Note that any time on ${\cal C}^-$ is
considered later than any time on ${\cal C}^+$. The source terms
in Eq.~(\ref{eq:definingZneq}) allow to generate correlation
functions by functional differentiation such as
\begin{align}
  \langle {\cal T_C}\Phi(x_1)\cdots\Phi(x_n)\rangle
  &= \left.\frac{\delta^nZ[J,K;\rho_D]}
                {i^n\delta J(x_1)\cdots\delta J(x_n)}
     \right|_{J,K\equiv0},
\label{eq:ClCorrFfromZ}
\end{align}
where the field indices have been suppressed and we have used that
for the closed time path $Z = 1$ in the absence of sources. We
have introduced, in Eq.~(\ref{eq:ClCorrFfromZ}), two contour
source terms, $J^{\cal C}$ and $K^{\cal C}$, which we use below to
go over by Legendre transformation to the corresponding 2PI
effective action.

\subsubsection{Functional integral}
For a theory with action (\ref{eq:Sq}) the generating functional
$Z[J,K;\rho_D]$ can be expressed in terms of a functional integral
using standard techniques (see e.g.\ Refs.~\cite{Chou1985a,Berges2005a} and
references therein):
\begin{align}
\label{eq:NEgenFuncZphipm}
  &Z[J,K;\rho_D]
  = \int [{\rm d}\varphi_0^+][{\rm d}\varphi_0^-]\,\,
  \rho_D\left[\varphi_0^+,\varphi_0^-\right]
  \nonumber\\
  &\times
     \int\limits_{\varphi^+(t_0,{\bf x})=\varphi_0^{+}({\bf x})}^{\varphi^-(t_0,{\bf x})
     =\varphi_0^{-}({\bf x})}{\cal D}^\prime\varphi^+{\cal D}^\prime\varphi^-
     \exp\Bigg\{i\Bigg[S[\varphi^+,\varphi^-]
  \nonumber\\
  &\ + \int_{x}\,\left( \varphi^+_a, \varphi^-_a\right)\left(\begin{array}{c} J^+_a \\ - J^-_a
  \end{array}\right)
  \nonumber\\
  &\ +\frac{1}{2}\int_{xy}\, \left( \varphi^+_a,\varphi^-_a
  \right)\left(\begin{array}{cc} K^{++}_{ab} & -K^{+-}_{ab}\\
  -K^{-+}_{ab} & K^{--}_{ab} \end{array}\right)
  \left(\begin{array}{c} \varphi^+_b \\ \varphi^-_b
  \end{array}\right)
\Bigg]\Bigg\},
\end{align}
where
$\rho_D(\varphi_0^+,\varphi_0^-)=\langle\varphi^+|\rho_D(t_0)|\varphi^-\rangle$
and the matrix elements are taken with respect to eigenstates of
the Heisenberg field operators at initial time, $\Phi_a(t_0,{\bf
x})|\varphi^{\pm} \rangle = \varphi_{0,a}^{\pm}({\bf x})
|\varphi^{\pm}\rangle$.
In Appendix \ref{app:inistate}, we provide an explicit expression for the initial-state density matrix $\rho_D(\varphi_0^+,\varphi_0^-)$ used later in our numerical calculations. 
The integral measures are given as
$[\mathrm{d}\varphi_0^{\pm}]=\prod_{a,{\bf x}}\mathrm{d}
\varphi^{\pm}_{0,a}({\bf
x})$ and ${\cal D}^\prime\varphi^{\pm} =
\prod_{a,x_0>t_0,{\bf x}}\mathrm{d} \varphi^{\pm}_a(x_0,{\bf x})$,
with the prime indicating that the integration over the fields at
initial time $t_0$ is excluded. The superscript `$+$' (`$-$')
indicates that the sources are taken to be different on the
forward (${\cal C}^+$) and backward (${\cal C}^-$) branch of the
closed time path. Because of the different sources, the
corresponding fields on the different branches are labelled
accordingly. The minus sign in front of the `$-$' terms accounts
for the reversed time integration. Using this notation the action
functional reads
\begin{align}
  S[\varphi^+,\varphi^-]
  &=
  \frac{1}{2}\int_{xy}\!\! \left( \varphi^+_a,\varphi^-_a \right)
  \left(\begin{array}{cc}
    iG^{-1}_{0,ab}   & 0\\
    0           &  -iG^{-1}_{0,ab}
  \end{array}\right)
  \left(\begin{array}{c} \varphi^+_b \\ \varphi^-_b
  \end{array}\right)
  \nonumber\\
  & -\frac{g}{8} \int_{x} \left(
   \varphi_a^+\varphi_a^+\varphi_b^+\varphi_b^+
  -\varphi_a^-\varphi_a^-\varphi_b^-\varphi_b^-\right) ,
\label{eq:Spm}
\end{align}
which corresponds to the defining action (\ref{eq:Sq}) if the time
integration is replaced by an integration along the closed time
contour ${\cal C}$.

In order to simplify the comparison with the classical statistical field theory below, a standard linear transformation $R$ of the fields is introduced as
\begin{align}
  \left(\begin{array}{c}
  \varphi_a \\
  {\tilde\varphi}_a
  \end{array}\right)
  &\equiv R\,\left(\begin{array}{c}
  \varphi_a^+ \\
  \varphi_a^-
  \end{array}\right),
\label{eq:RTransf}
\end{align}
where
\begin{align}
  &R
  = \left(\begin{array}{rr} \frac{1}{2} & \frac{1}{2} \\
                             1           & -1
                             \end{array}\right), \qquad
  R^{-1}
  = \left(\begin{array}{rr} 1 & \frac{1}{2} \\
                             1 & -\frac{1}{2}
                             \end{array}\right)
\label{eq:RMatrix}
\end{align}
such that $\varphi_a = (\varphi^+_a + \varphi^-_a)/2$ and ${\tilde
\varphi}_a = \varphi^+_a - \varphi^-_a$, or, $\varphi^+_a =
\varphi_a + {\tilde \varphi}_a/2$ and $\varphi^-_a = \varphi_a -
{\tilde \varphi}_a/2$, respectively. To avoid a proliferation of
symbols we have used here $\varphi_a$, which agrees with the
defining field in (\ref{eq:Sq}) only for $\varphi^+_a =
\varphi^-_a$. Since this will be the case for expectation values
in the absence of sources, where physical observables are
obtained, and since there is no danger of confusion in the
following we keep this notation.

Correspondingly, we write for the source terms
\begin{eqnarray}
  \left(\begin{array}{c} J_a \\ \tilde{J}_a \end{array}\right)
  &\equiv& R\,\left(\begin{array}{c} J_a^+ \\ J_a^-
  \end{array}\right) ,
\label{eq:JTransf} \\
 \left(\begin{array}{rr}
  K^F_{ab} & K^{\mathrm R}_{ab} \\
  K^{\mathrm A}_{ab} & K^{\tilde F}_{ab}
  \end{array}\right)
  &\equiv& R
  \left(\begin{array}{rr}
  K^{++}_{ab}  & K^{+-}_{ab} \\
  K^{-+}_{ab}  & K^{--}_{ab}
  \end{array}\right)R^{T} .
\label{eq:KMatrix}
\end{eqnarray}
Inserting these definitions into the functional integral
(\ref{eq:NEgenFuncZphipm}) and using that (\ref{eq:JTransf}) and
(\ref{eq:KMatrix}) can be equivalently written as
\begin{eqnarray}
  \left(\begin{array}{c}
  \tilde{J}_a \\ J_a
  \end{array}
  \right)
  &\equiv& \left(R^{-1}\right)^T
  \left(\begin{array}{c}
  J_a^+ \\ - J_a^-
  \end{array}\right),
\label{eq:JTransf2}\\
 \left(\begin{array}{rr}
  K^{\tilde F}_{ab}     & K^{\mathrm A}_{ab} \\
  K^{\mathrm R}_{ab}      & K^F_{ab}
  \end{array}\right)
  &\equiv& \left(R^{-1}\right)^T
  \left(\begin{array}{rr}
  K^{++}_{ab}  & -K^{+-}_{ab} \\
  -K^{-+}_{ab} & K^{--}_{ab}
  \end{array}\right)R^{-1}\qquad
\label{eq:KMatrix2}
\end{eqnarray}
one finds:
\begin{eqnarray}
  && Z[J,\tilde{J},K^F,K^\mathrm{R},K^\mathrm{A},K^{\tilde
  F};\rho_D]\nonumber\\[0.4ex]
  && = \int [{\rm d}\varphi_0][{\rm d}{\tilde\varphi}_0]\,
       \rho_D\left[\varphi_0+{\tilde\varphi}_0/2,
                   \varphi_0-{\tilde\varphi}_0/2\right]
  \nonumber\\
  && \times
     \int\limits_{\varphi_0,{\tilde\varphi}_0}
        {\cal D}^\prime\varphi{\cal D}^\prime{\tilde\varphi}
     \,\exp\Bigg\{i\Bigg[S[\varphi,{\tilde\varphi}]
     + \int_x\,\left( \varphi_a, {\tilde\varphi}_a\right)
               \left(\begin{array}{c}{\tilde J}_a \\ J_a\end{array}\right)
  \nonumber\\
  && +\frac{1}{2}\int_{xy}\,
  \left( \varphi_a,{\tilde\varphi}_a \right)
  \left(\begin{array}{cc}
  K^{\tilde F}_{ab}         & K^\mathrm{A}_{ab}\\[1pt]
  K^\mathrm{R}_{ab}         & K^{F}_{ab}
  \end{array}\right)
  \left(\begin{array}{c} \varphi_b \\ {\tilde\varphi}_b\end{array}\right)
  \Bigg]\Bigg\},
\label{eq:ZQuantPhiXi}
\end{eqnarray}
where $S[\varphi,{\tilde\varphi}] = S_0[\varphi,{\tilde\varphi}] +
S_{\rm int}[\varphi,{\tilde\varphi}]$ consists of the action for
the free field theory
\begin{equation}
  S_0[\varphi,{\tilde\varphi}]
  = \frac{1}{2}\int_{xy}\,
  \left( \varphi_a,{\tilde\varphi}_a \right)
  \left(\begin{array}{cc}
  0         & iG^{-1}_{0,ab}\\
  iG^{-1}_{0,ab}    & 0
  \end{array}\right)
  \left(\begin{array}{c} \varphi_b \\ {\tilde\varphi}_b \end{array}\right)
  \label{eq:freeS}
\end{equation}
and the interaction part
\begin{eqnarray}
  S_{\rm int}[\varphi,{\tilde\varphi}]
  &=& -\frac{g}{2}\int_x {\tilde\varphi}_a(x)\varphi_a(x)\varphi_b(x)\varphi_b(x)
  \nonumber\\
  &&  -\frac{g}{8}\int_x {\tilde\varphi}_a(x){\tilde\varphi}_a(x)
                         {\tilde\varphi}_b(x)\varphi_b(x).
\label{eq:SqPhiXi}
\end{eqnarray}
%

\subsubsection{Connected one- and two-point functions}
\label{sec:QCorrFcts}
From the generating functional for connected correlation functions
\begin{align}
  W = -i\, {\rm ln} Z
  \label{eq:WGenF}
\end{align}
we define the macroscopic field $\phi_a$, and ${\tilde\phi}_a$ by
\begin{align}
  \frac{\delta W}{\delta {\tilde J}_a(x)} = \phi_a(x) , \quad
  \frac{\delta W}{\delta J_a(x)} = {\tilde\phi}_a(x)\, .
\label{eq:Defphiphitilde}
\end{align}
The connected statistical correlation function $F_{ab}(x,y)$, the
retarded/advanced propagators $G^\mathrm{R/A}_{ab}(x,y)$, and
${\tilde F}_{ab}(x,y)$ are defined by
\begin{eqnarray}
  \frac{\delta W}{\delta K^{\tilde F}_{ab}(x,y)}
  &=&
  \frac{1}{2}\left( \phi_a(x)\phi_b(y) + F_{ab}(x,y) \right) \, ,
  \nonumber\\
  \frac{\delta W}{\delta K^\mathrm{A}_{ab}(x,y)}
  &=& \frac{1}{2}\left(\phi_a(x){\tilde\phi}_b(y)-iG^\mathrm{R}_{ab}(x,y) \right) \, ,
  \nonumber\\
  \frac{\delta W}{\delta K^\mathrm{R}_{ab}(x,y)}
  &=& \frac{1}{2}\left({\tilde\phi}_a(x)\phi_b(y)-iG^\mathrm{A}_{ab}(x,y) \right) \, ,
  \nonumber\\
  \frac{\delta W}{\delta K^{F}_{ab}(x,y)}
  &=& \frac{1}{2}\left({\tilde \phi}_a(x) {\tilde \phi}_b(y)
                     + {\tilde F}_{ab}(x,y)\right) \, .
  \label{eq:propagators}
\end{eqnarray}

For vanishing sources the field $\tilde\phi\equiv0$ and the propagator $\tilde
F\equiv0$ \cite{Chou1985a}. Moreover, the retarded and advanced correlators
$G^\mathrm{R}(x,y)$ and $G^\mathrm{A}(y,x)$ vanish for
$x_0<y_0$. (See also the discussion in Sect.~\ref{sect:classcorr}.) 
In the absence of sources, these two-point functions
are related then through the transformation (\ref{eq:RMatrix}),
\begin{eqnarray}
\left(\begin{array}{cc}
    F_{ab} &
    -i G^\mathrm{R}_{ab} \\
    -i G^\mathrm{A}_{ab} &
    0 \end{array}\right)
    = R
   \left(\begin{array}{cc}
   G_{ab}^{++} & G_{ab}^{+-}\\
   G_{ab}^{-+} & G_{ab}^{--}
   \end{array}\right)
   R^{T},
   \label{eq:GR}
\end{eqnarray}
to the time-ordered correlation functions written in the `$\pm$'
basis. The inverse of the two-point function matrix (\ref{eq:GR})
reads
\begin{equation}
\label{eq:inverse}
   \left(\begin{array}{cc}
   0            &  i(G^\mathrm{A})^{-1}_{ab} \\
   i(G^\mathrm{R})^{-1}_{ab} &  X^{-1}_{ab}
   \end{array}\right),
\end{equation}
where
\begin{align}
  X^{-1}_{ab}(x,y)
  &\equiv \int_{z w}
  (G^\mathrm{R})^{-1}_{ac}(x,z)
  F_{cd}(z,w)
  (G^\mathrm{A})^{-1}_{db}(w,y)
  \nonumber\\
  &\equiv \left[(G^\mathrm{R})^{-1} \cdot F \cdot (G^\mathrm{A})^{-1}
  \right]_{ab}(x,y).
  \label{eq:Xinv}
\end{align}
The last equation introduces a compact matrix notation that will
be employed below.
\\

\subsubsection{2PI effective action}
\label{sec:EOMClvsQ}

\label{app:2PIEA}

\begin{figure}[tb]
\begin{center}
\includegraphics[width=0.45\textwidth]{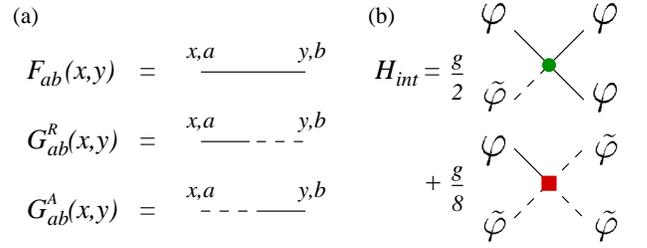}
\end{center}
\vspace*{-3ex} \caption{(color online) (a) Diagrammatic
representation of the correlators in the $\varphi$-$\tilde\varphi$
basis. A full line indicates the $1$- or $\varphi$-component, a
broken line the $2$- or $\tilde\varphi$-component. $F_{ab}(x,y)$
is the statistical correlation function,
$G^R_{ab}(x,y)=\rho_{ab}(x,y)\theta(x_0-y_0)$ and
$G^A_{ab}(x,y)=-\rho_{ab}(x,y)\theta(y_0-x_0)$ the retarded and
advanced Green's functions, respectively. Their representation in
terms of the real-valued spectral correlation function
$\rho_{ab}(x,y)$ exposes the $\theta$-functions which imply the
respective time ordering in $x_0$, $y_0$. (b) Diagrammatic
expansion of the quantum vertex term entering the action
(\ref{eq:Sq}). The classical action (\ref{eq:SqPhiXicl}) is
lacking the second contribution (red square).}
\label{fig:DiagramsFGRGAHI}
\end{figure}

The 2PI effective action is obtained as the Legendre transform
\begin{align}
  \Gamma
  =&\ W - \int_x \left(\phi_a(x) {\tilde J}_a(x)
     + {\tilde\phi}_a(x) J_a(x) \right)
  \nonumber\\
  & -\ \frac{1}{2} \int_{xy} \Big\{ K^{\tilde F}_{ab}(x,y)
    \Big(\phi_a(x)\phi_b(y) + F_{ab}(x,y) \Big)
  \nonumber\\
  &\qquad +\ K^\mathrm{A}_{ab}(x,y)
  \left(\phi_a(x){\tilde \phi}_b(y) - iG^\mathrm{R}_{ab}(x,y) \right)
  \nonumber\\
  &\qquad +\ K^\mathrm{R}_{ab}(x,y)
  \left({\tilde \phi}_a(x)\phi_b(y) -iG^\mathrm{A}_{ab}(x,y)\right)
  \nonumber\\
  &\qquad +\ K^{F}_{ab}(x,y)
  \left( {\tilde \phi}_a(x){\tilde \phi}_b(y) + {\tilde F}_{ab}(x,y) \right)
  \Big\} .
  \label{eq:effact}
\end{align}
This corresponds to the 2PI effective action originally discussed
in Refs.~\cite{Luttinger1960a,Baym1962a,Cornwall1974a}, however,
written in terms of the rotated variables. From (\ref{eq:effact})
one observes the equations of motion for the fields
\begin{align}
  \frac{\delta \Gamma}{\delta \phi_a(x)}
  =&\ - {\tilde J}_a(x) -\int_y
      \Big( K^{\tilde F}_{ab}(x,y) \phi_b(y)
  \nonumber\\
  & +\ \frac{1}{2} K^\mathrm{A}_{ab}(x,y) {\tilde \phi}_b(y)
    +  \frac{1}{2}{\tilde \phi}_b(y) K^\mathrm{R}_{ba}(y,x) \Big),
  \\
  \frac{\delta \Gamma}{\delta \tilde{\phi}_a(x)}
  =&\ - J_a(x)
     -\int_y \Big( K^{F}_{ab}(x,y) \tilde{\phi}_b(y)
  \nonumber\\
  & +\ \frac{1}{2} K^\mathrm{R}_{ab}(x,y) \phi_b(y)
    +  \frac{1}{2} \phi_b(y) K^\mathrm{A}_{ba}(y,x) \Big),
    \label{eq:fields}
\end{align}
as well as for the two-point functions
\begin{align}
  \frac{\delta \Gamma}{\delta F_{ab}(x,y)}
  &= - \frac{1}{2}K^{\tilde F}_{ab}(x,y) ,
  \\
  i\frac{\delta \Gamma}{\delta G^\mathrm{R}_{ab}(x,y)}
  &= - \frac{1}{2}K^\mathrm{A}_{ab}(x,y) ,
  \\
  i\frac{\delta \Gamma}{\delta G^\mathrm{A}_{ab}(x,y)}
  &= - \frac{1}{2}K^\mathrm{R}_{ab}(x,y) ,
  \\
  \frac{\delta \Gamma}{\delta \tilde{F}_{ab}(x,y)}
  &= - \frac{1}{2}K^{F}_{ab}(x,y) .
\end{align}
The diagrammatic calculation of the 2PI effective action involves
all closed two-particle irreducible Feynman graphs with lines
associated to the full two-point
correlators~\cite{Luttinger1960a,Baym1962a,Cornwall1974a}. This is
illustrated in Fig.~\ref{fig:DiagramsFGRGAHI} for vanishing
macroscopic field $\phi$ and in the absence of external sources.

In the presence of a non-vanishing field value, $\phi\not = 0$,
the interaction vertices are obtained from $(\ref{eq:SqPhiXi})$ by
shifting in $S[\varphi, {\tilde \varphi}]$ the field $\varphi \to
\phi + \varphi$, and collecting all cubic and quartic terms in the
fluctuating fields $\varphi$ and ${\tilde\varphi}$, i.e.\
\begin{eqnarray}
  S_{\rm int}[\varphi,{\tilde \varphi};\phi]
  &=&
      -\frac{g}{2}\int_x
  {\tilde\varphi}_a(x)        \varphi_a(x)        \varphi_b(x)\varphi_b(x)
  \nonumber\\
  &&  -\frac{g}{8}\int_x
  {\tilde\varphi}_a(x){\tilde\varphi}_a(x){\tilde\varphi}_b(x)\varphi_b(x)
  \nonumber\\
  &&  - g         \int_x
  {\tilde\varphi}_a(x)        \varphi_a(x)        \varphi_b(x)   \phi_b(x)
  \nonumber\\
  &&  -\frac{g}{2}\int_x
  {\tilde\varphi}_a(x)           \phi_a(x)        \varphi_b(x)\varphi_b(x)
  \nonumber\\
  &&  -\frac{g}{8}\int_x
  {\tilde\varphi}_a(x){\tilde\varphi}_a(x){\tilde\varphi}_b(x)   \phi_b(x).
\label{eq:SqPhiXiVev}
\end{eqnarray}
The quadratic terms in the fluctuating fields are taken into
account in the classical inverse propagator (\ref{eq:invfreeProp})
by the replacement
\begin{eqnarray}
  H_\mathrm{1B}\delta_{ab} \to \left[ H_\mathrm{1B} + \frac{g}{2}
  \phi_c(x) \phi_c(x)\right] \delta_{ab} + g \phi_a(x) \phi_b(x) .
  \label{eq:H1Bwithmeanfield}
\end{eqnarray}
In the free part of the action (\ref{eq:freeS}), and in the
dynamic equations derived below, this corresponds to a field
dependent $iG_0^{-1}(x,y;\phi)$, while the general form of the
equations remains unchanged \footnote{%
In Ref.~\cite{Gasenzer2005a} we used the notation
$iG_0^{-1}(\phi\equiv0)=iD^{-1}$.}. We note that linear terms in
the fluctuating fields ensure cancellation of possible tadpole
contributions, which therefore do need not to be considered
explicitly.

\subsubsection{Exact evolution equations}

For vanishing sources, the exact inverse two-point function
(\ref{eq:inverse}) can then be written
as~\cite{Luttinger1960a,Baym1962a,Cornwall1974a}
\begin{eqnarray}
\label{eq:DSeqn}
   &&\left(\begin{array}{cc}
   0            & i(G^\mathrm{A})^{-1}_{ab} \\
   i(G^\mathrm{R})^{-1}_{ab} & X^{-1}_{ab}
   \end{array}\right)
   \nonumber\\
   &&\qquad =\  \left(\begin{array}{cc}
   0        & G_{0,ab}^{-1} \\
   G_{0,ab}^{-1}    & 0
   \end{array}\right)
   -  \left(\begin{array}{cc}
    0           & -i\Sigma^\mathrm{A}_{ab}\\
    -i \Sigma^\mathrm{R}_{ab} & \Sigma^F_{ab}
    \end{array}\right),\qquad
\end{eqnarray}
where $X^{-1}$ is defined in Eq.~(\ref{eq:Xinv}) and the retarded,
advanced and statistical self-energies are related through the
transformation (\ref{eq:RMatrix}),
\begin{equation}
\left(\begin{array}{cc}
    0 &
    -i \Sigma^\mathrm{A}_{ab} \\
    -i \Sigma^\mathrm{R}_{ab} &
    \Sigma^F_{ab} \end{array}\right)
    = \left(R^{-1}\right)^T
   \left(\begin{array}{cc}
   \Sigma_{ab}^{++} & -\Sigma_{ab}^{+-}\\
   -\Sigma_{ab}^{-+} & \Sigma_{ab}^{--}
   \end{array}\right)
   R^{-1},
   \label{eq:Sigmarotated}
\end{equation}
to the self-energies written in the `$\pm$' basis. To the
retarded/advanced self-energy $\Sigma^\mathrm{R/A}$ and the
statistical self-energy $\Sigma^F$ contribute only graphs with
propagator lines associated to $G^\mathrm{R,A}$ and $F$, which can
be obtained from closed two-particle irreducible graphs by opening
one propagator line~\cite{Luttinger1960a,Baym1962a,Cornwall1974a}.

To convert Eq.~(\ref{eq:DSeqn}) for the inverse propagator into an
equation, which is more suitable for initial value problems, we
convolute with the propagator matrix (\ref{eq:GR}) from the right
and with the classical propagator from the left. This yields the
Schwinger-Dyson equations for the retarded/advanced propagator in
the absence of sources,
\begin{equation}
  G^\mathrm{R/A}
  = G^\mathrm{R/A}_0 - G^\mathrm{R/A}_0 \cdot \Sigma^\mathrm{R/A} \cdot
  G^\mathrm{R/A},
  \label{eq:RA}
\end{equation}
and the statistical propagator,
\begin{equation}
  F = F_0 - F_0 \cdot \Sigma^\mathrm{A} \cdot G^\mathrm{A}
          - G^\mathrm{R}_0 \cdot \left[\Sigma^\mathrm{R} \cdot F
                                 + \Sigma^F \cdot G^\mathrm{A} \right]
                                 ,
  \label{eq:SDF}
\end{equation}
where we have used the compact notation introduced in
Eq.~(\ref{eq:Xinv}). For the spectral function,
\begin{align}
  \rho_{ab}(x,y)
  &= G_{ab}^\mathrm{R}(x,y)-G_{ab}^\mathrm{A}(x,y) ,
  \label{eq:rhoinGRA}
\end{align}
the Schwinger-Dyson equation follows from Eqs.~(\ref{eq:rhoinGRA})
and (\ref{eq:RA}) as
\begin{eqnarray}
  \rho
  &=& \rho_0 - G^\mathrm{R}_0 \cdot \Sigma^\mathrm{R} \cdot G^\mathrm{R}
         + G^\mathrm{A}_0 \cdot \Sigma^\mathrm{A} \cdot G^\mathrm{A}.
  \label{eq:SDrho}
\end{eqnarray}
Acting on Eqs.~(\ref{eq:SDF}) and (\ref{eq:SDrho}) with $G_0^{-1}$
from the left brings these equations in a form which is more
suitable for initial-value problems. For this we write
\begin{equation}
  \Sigma_{ab}^\mathrm{R}(x,y;\phi) = \Sigma^{(0)}_{ab}(x) \delta(x-y)
  + \theta(x_0-y_0)\Sigma^{\rho}_{ab}(x,y;\phi)  ,
\end{equation}
where we introduce the spectral part of the self-energy
\begin{equation}
  \Sigma^{\rho}_{ab}(x,y;\phi)
  = \Sigma_{ab}^\mathrm{R}(x,y;\phi)-\Sigma_{ab}^\mathrm{A}(x,y;\phi) .
  \label{eq:sigmainGRA}
\end{equation}
Using that $F_0$ and $\rho_0$ satisfy the homogeneous field
equations
\begin{eqnarray}
  \int_z G_{0,ac}^{-1}(x,z)F_{0,cb}(z,y)&=&0\, ,
  \nonumber\\
  \int_z G_{0,ac}^{-1}(x,z)\rho_{0,cb}(z,y)&=&0\,,
  \label{eq:F0rho0}
\end{eqnarray}
and taking into account all $\theta$-functions one obtains the
dynamic equations for the two-point correlation functions $F$ and
$\rho$ \cite{Gasenzer2005a,Aarts2001a,Aarts2002a}:
\begin{eqnarray}
&&\left[i\sigma^2_{ac}\partial_{x_0} + M_{ac}(x) \right]
F_{cb}(x,y)
 \nonumber\\
&&\qquad\qquad = - \int_{t_0}^{x_0} \! {\rm d} z\,
 \Sigma^{\rho}_{ac}(x,z;\phi) F_{cb}(z,y)
\nonumber \\
&&\qquad\qquad\quad + \int_{t_0}^{y_0} \! {\rm d}z\,
\Sigma^{F}_{ac}(x,z;\phi) \rho_{cb}(z,y) ,
\nonumber\\[1.5ex]
&&\left[i\sigma^2_{ac}\partial_{x_0}
    +M_{ac}(x) \right] \rho_{cb} (x,y)  \nonumber\\
&&\qquad\qquad = -\int_{y_0}^{x_0} \! {\rm d}z\,
\Sigma^{\rho}_{ac} (x,z;\phi)
 \rho_{cb}(z,y), \qquad
 \label{eq:EOMFrho}
\end{eqnarray}
where we employ the notation
\begin{equation}
 \int^{x_0}_{t_0} \!{\rm d}z 
 \equiv \int^{x_0}_{t_0} {\rm d}z_0 \int {\rm d}^d z.
\end{equation}
Here
\begin{eqnarray}
\label{eq:MofF}
  M_{ab}(x)
   &=& \delta_{ab}
  \Big[H_\mathrm{1B}(x)
   + \frac{g}{2}\Big(\phi_c(x)\phi_c(x)+F_{cc}(x,x)\Big)\Big]
  \nonumber\\
  &&
   + g\Big(\phi_a(x)\phi_b(x)+F_{ab}(x,x)\Big)
\end{eqnarray}
is the mean-field energy matrix which includes the field-dependent
terms of the classical inverse propagator $iG_{0,ab}^{-1}(\phi)$
defined in Eqs.~(\ref{eq:invfreeProp}) with
(\ref{eq:H1Bwithmeanfield}), and the local part of the
self-energy,
\begin{equation}
\Sigma_{ab}^{(0)}(x) = \frac{g}{2} \delta_{ab} F_{cc}(x,x) + g
F_{ab}(x,x) ,
\end{equation}
as defined in Eq.~(\ref{eq:sigmainGRA}). We note that the spectral function
at initial time is characterized by the commutator (\ref{eq:BoseCommutator}),
with
\begin{align}
   \left.\rho_{ab}(x,y)\right|_{x_0=y_0}= -i\sigma^2_{ab}\delta({\bf x}-{\bf y}),
   \label{eq:equaltimerho}
\end{align}
for the non-relativistic theory.

The dynamic equation for the mean field $\phi_a(x)$ is obtained
from Eq.~(\ref{eq:fields}) with $\tilde{\phi}_a(x) \equiv 0$ in
the absence of sources. It reads \cite{Gasenzer2005a,Aarts2002b}:
\begin{eqnarray}
\label{eq:EOMphi}
&&   \Big(-i\sigma^2_{ab}\partial_{x_0}
    -g\,F_{ab}(x,x)\Big)\phi_b(x) -\Big(H_\mathrm{1B}(x)
    \nonumber\\
&&  +\frac{g}{2}\big[\phi_c(x)\phi_c(x)
    +F_{cc}(x,x)\big]\Big)
    \phi_a(x) = \frac{\delta \Gamma_2}{\delta\phi_a(x)} \, , \quad
\end{eqnarray}
where the functional derivative is taken for fixed two-point
functions and $\Gamma_2$ contains all closed 2PI
graphs~\cite{Luttinger1960a,Baym1962a,Cornwall1974a}. In
Sec.~\ref{sec:SelfEnClvsQ} we will employ an expansion in the
number of field components to next-to-leading order to
approximately describe the dynamics, for which one obtains the
compact expression~\cite{Aarts2002b}
\begin{eqnarray}
   \frac{\delta \Gamma_2}{\delta\phi_a(x)} &=&
   \int_{t_0}^{x_0} \!{\rm d}y\,
   \Sigma^\rho_{ab}(x,y;\phi\equiv 0)\,\phi_b(y)
   \, .
\end{eqnarray}

\subsection{Classical statistical dynamics}
\label{sec:FIntCl}
%
\subsubsection{Classical equation of motion}
The classical field equation of motion can be derived from the
defining action (\ref{eq:Sq}). In the basis (\ref{eq:RTransf}) the
same equation of motion for the field $\varphi_a(x)$ is obtained
from the action with (\ref{eq:freeS}) and (\ref{eq:SqPhiXi}) by
\begin{equation}
  \frac{\delta S[\varphi,{\tilde \varphi}]}
       {\delta {\tilde\varphi}}\Big|_{\varphi=\varphi^\mathrm{cl},{\tilde \varphi}=0}
= \frac{\delta S^\mathrm{cl}[\varphi,{\tilde \varphi}]}
       {\delta {\tilde \varphi}}\Big|_{\varphi = \varphi^\mathrm{cl}} =
       0,
\label{eq:classicaleom}
\end{equation}
as a consequence of setting ${\tilde \varphi}=0$. For the first
identity in (\ref{eq:classicaleom}) we have defined
$S^\mathrm{cl}[\varphi,{\tilde \varphi}] \equiv
S_0[\varphi,{\tilde \varphi}] + S_{\rm
int}^\mathrm{cl}[\varphi,{\tilde \varphi}]$ as that part of
$S[\varphi,{\tilde\varphi}]$ which is linear in ${\tilde\varphi}$.
As a consequence, the non-interacting part
$S_0[\varphi,{\tilde\varphi}]$ agrees with the respective
expression appearing for the quantum theory $(\ref{eq:freeS})$.
The interaction part of $S^\mathrm{cl}[\varphi,{\tilde \varphi}]$,
however, differs from (\ref{eq:SqPhiXi}) in that it contains fewer
vertices:
\begin{equation}
  S_{\rm int}^\mathrm{cl}[\varphi,{\tilde \varphi}]
  = -\frac{g}{2}\int_x {\tilde\varphi}_a(x)\varphi_a(x)\varphi_b(x)\varphi_b(x).
  \label{eq:SqPhiXicl}
\end{equation}
This is illustrated in Fig.~\ref{fig:DiagramsFGRGAHI}b. The
absence of vertices beyond those which are linear in
${\tilde\varphi}$ turns out to be the crucial difference between a
classical and a quantum statistical field theory as is discussed
in the following.

\subsubsection{Classical statistical generating functional}
We will construct the generating functional for the classical
statistical field theory, $Z^\mathrm{cl}[J,K;W]$ for given
probability functional $W$ for the fields at initial time, similar
to the expression (\ref{eq:ZQuantPhiXi}) for the quantum theory.
For this, we rewrite the equation of motion
(\ref{eq:classicaleom}) as a $\delta$-constraint in a functional
integral using the Fourier transform representation
\begin{eqnarray}
  &&\delta\left[\frac{\delta S^\mathrm{cl}[\varphi,{\tilde\varphi}]}
                   {\delta {\tilde \varphi}}\right]
  = \int {\cal D} {\tilde\varphi}\,
    \exp\left\{i \int_x \frac{\delta S^\mathrm{cl}[\varphi,{\tilde\varphi}]}
                   {\delta {\tilde \varphi_a(x)}} \, {\tilde
                   \varphi_a(x)}\right\}
  \nonumber\\
  && = \int {\cal D} {\tilde\varphi}\,
    \exp\left\{i S^\mathrm{cl}[\varphi,{\tilde \varphi}]+{i}
    \int_{\mathbf{x}}\pi_{0,a}({\bf x})\tilde\varphi_{0,a}({\bf x})\right\}
\label{eq:deltaconstraint}.
\end{eqnarray}
For the last equality, the second term in the exponent subtracts
the boundary term at the initial time $t_0$, which follows from
partial integration. We emphasize that we have to take into
account this boundary term since $t_0$ will be taken to be finite
for the nonequilibrium initial-value problems considered. 
Note that in the path integral (\ref{eq:ZQuantPhiXi}), the action $S[\varphi,\tilde\varphi]$ is implied not to depend on boundary terms at initial time $t_0$. 
(As a consequence, the term $i\int_{\mathbf{x}}\pi_{0,a}({\bf x})\tilde\varphi_{0,a}({\bf x})$ appears in Eq.~(\ref{eq:deltaconstraint}) and not half of it.)
Here
$\pi_{0,a}({\bf x})=\pi_a(x_0=t_0,{\bf x})$ is the initial
canonical momentum, with $\pi_a(x) = i\sigma^2_{ab}\varphi_b(x)$
for the non-relativistic theory as in Eq.~(\ref{eq:canmom}).
We note that Eq.~(\ref{eq:deltaconstraint}) remains valid also for 
a relativistic theory, where $\pi_{0,a}({\bf x})$ denotes the
time derivative of the field at time $t_0$.

It should be stressed that the Fourier transform expression
(\ref{eq:deltaconstraint}) could not be achieved with
$S^\mathrm{cl}[\varphi,{\tilde \varphi}]$ replaced by
$S[\varphi,{\tilde\varphi}]$ since the latter is not linear in
${\tilde \varphi}$. For the non-interacting theory, however,
$S_0^\mathrm{cl}[\varphi,{\tilde \varphi}] = S_{0}[\varphi,{\tilde
\varphi}]$ holds, and we will recover the fact that the free
classical and quantum theories are governed by the same dynamics
for same initial conditions. The same fact holds true for all
Gaussian (leading-order large-$\cal N$ or Hartree-type)
approximations. Therefore, it is crucial for a quantum classical
comparison that we go beyond and consider the next-to-leading
order corrections as we do in Sect.~\ref{sec:EOMClvsQ} below.

Any functional $f[\varphi^\mathrm{cl}]$ of the classical field
$\varphi^\mathrm{cl}$, being a solution of the classical equation
of motion (\ref{eq:classicaleom}), can be written as
\begin{align}
  f[\varphi^\mathrm{cl}]
  &= \int\limits_{\varphi_0=\varphi_0^\mathrm{cl}}\!\! {\cal D}^\prime \varphi \, f[\varphi]\,
      \delta\left[\varphi^\mathrm{cl} - \varphi\right]
  \nonumber\\
  &= \int\limits_{\varphi_0=\varphi_0^\mathrm{cl}}\!\! {\cal D}^\prime \varphi\, f[\varphi]\,
      \delta\left[\frac{\delta S^\mathrm{cl}[\varphi,{\tilde\varphi}]}
                       {\delta {\tilde \varphi}}\right] {\cal J}[\varphi]
  \nonumber\\
  &= \int\limits_{\varphi_0=\varphi_0^\mathrm{cl}}\!\!
     {\cal D}'\varphi\, {\cal D}{\tilde \varphi}\,
      f[\varphi]\,
  \nonumber\\
  &   \times\ \exp\Big\{i S^\mathrm{cl}[\varphi,{\tilde \varphi}]
      + {i}\int_{\mathbf{x}}\pi_{0,a}\tilde\varphi_{0,a}\Big\}
      {\cal J}[\varphi],
  \label{eq:functionalasPI}
\end{align}
where the Jacobian reads
\begin{equation}
  {\cal J}[\varphi]
  = \left|{\rm det}\left(
  \frac{\delta^2 S^\mathrm{cl}[\varphi,{\tilde \varphi}]}
       {\delta \varphi \delta {\tilde\varphi}}\right)\right|.
  \label{eq:Jacobean}
\end{equation}
Here the Jacobian plays the role of an irrelevant normalization constant, 
which has been discussed in detail in Ref.~\cite{Jeon2005a} and references
therein.

Classical correlation functions are obtained as phase-space
averages over trajectories given by solutions of the classical
field equation $(\ref{eq:classicaleom})$. Such averages, for an
arbitrary functional of the field, are defined as
\begin{equation}
  \langle f[\varphi_a]\rangle_\mathrm{cl}
  = \int[\mathrm{d}\varphi_0^\mathrm{cl}][\mathrm{d}\pi_0^\mathrm{cl}]  
     W[\varphi_0^\mathrm{cl},\pi_0^\mathrm{cl}]
    f[\varphi_a^\mathrm{cl}].
  \label{eq:classaverage}
\end{equation}
Here $W[\varphi_0^\mathrm{cl},\pi_0^\mathrm{cl}]$ denotes the normalized phase-space probability functional at initial time (and is not to be confused with the generating functional (\ref{eq:WGenF})). 
In Appendix \ref{app:inistate} we provide an explicit expression for $W[\varphi_0^\mathrm{cl},\pi_0^\mathrm{cl}]$, a functional of four fields which is symmetric under the exchange of $\pi_{0,a}^\mathrm{cl}$ and $i\sigma^2_{ab}\varphi_{0,b}^\mathrm{cl}$ for $a=1,2$, since the canonical momentum (\ref{eq:canmom}) is proportional to the field itself.
The measure indicates integration over the classical phase-space. 
The theory may be defined on a spatial lattice to regulate the Rayleigh-Jeans divergence of classical statistical field theory.

Using Eqs.~(\ref{eq:functionalasPI}) and (\ref{eq:classaverage}),
we can now write down a generating functional for classical
correlation functions in the form:
\begin{align}
  & Z^\mathrm{cl}[J,\tilde{J},K^F,K^R,K^A,K^{\tilde F};W]
  = \int [{\rm d}\varphi_0^\mathrm{cl}][\mathrm{d}\pi_0^\mathrm{cl}]  
     W[\varphi_0^\mathrm{cl},\pi_0^\mathrm{cl}]
  \nonumber\\
  &\times
     \int\limits_{{\varphi_0=\varphi_0^\mathrm{cl}, }
                           {\pi_0=\pi_0^\mathrm{cl}}}
     \!\!\!\!\!\!{\cal D}'\varphi{\cal D}{\tilde\varphi}
     \,\exp\Bigg\{i\Bigg[S^\mathrm{cl}[\varphi,{\tilde \varphi}]
     +\int_{\mathbf{x}}\pi_{0,a}\tilde\varphi_{0,a}
  \nonumber\\
  &\qquad
  +\ \int_x\,\left( \varphi_a, {\tilde \varphi}_a\right)
            \left(\begin{array}{c}{\tilde J}_a \\ J_a\end{array}\right)
  \nonumber\\
  &\qquad +\ \frac{1}{2}\int_{xy}\,
  \left( \varphi_a,{\tilde \varphi}_a\right)
  \left(\begin{array}{cc}
    K^{\tilde F}_{ab}   & K^{A}_{ab}\\[1pt]
    K^{R}_{ab}      & K^{F}_{ab}
  \end{array}\right)
  \left(\begin{array}{c}
    \varphi_b \\ {\tilde \varphi}_b
  \end{array}\right)
  \Bigg]\Bigg\} {\cal J}[\varphi],
\label{eq:ZClassPhiXi}
\end{align}
where $S^\mathrm{cl}[\varphi,{\tilde \varphi}]=S_0[\varphi,{\tilde
\varphi}]+S_{\rm int}^\mathrm{cl}[\varphi,{\tilde \varphi}]$ as
defined in Eqs.~(\ref{eq:freeS}) and (\ref{eq:SqPhiXicl}), and where the relation (\ref{eq:canmom}) between the field and the canonical momentum is implied.

Comparing with the quantum generating functional in
Eq.~(\ref{eq:ZQuantPhiXi}), and using that the Jacobian ${\cal
J}[\varphi]$ plays the role of an irrelevant normalization constant, 
we find that the classical functional
(\ref{eq:ZClassPhiXi}) takes the same Lagrangian form, with
initial conditions given by a density matrix $\rho_D$ which is
characterized by the Fourier transform of the phase-space probability distribution $W[\varphi_0,\pi_0]$:
\begin{align}
  &\rho_D
    \left[\varphi_0+{\tilde \varphi}_0/2,\varphi_0-{\tilde \varphi}_0/2\right]  
  \nonumber\\
  &\qquad
  =\ \int[\mathrm{d}\pi_0]
  W[\varphi_0,\pi_0]\exp\left\{i\int_{\mathbf{x}}
                    \pi_{0,a}\,\tilde\varphi_{0,a}\right\}.
  \label{eq:FTrhoW}
\end{align}

In summary, the generating functionals for correlation functions are very similar in the quantum and the classical statistical theory. 
The crucial
difference is that the quantum theory is characterized by more
vertices. As a consequence, we can follow the very same steps as
in Sect.~\ref{sec:FIntQ} to construct the classical statistical
2PI effective action and to derive the time evolution equations
from it.

Similar to the discussion for the quantum theory above, for
$\phi\not = 0$ the interaction vertices for the classical
statistical theory are obtained by shifting in
$S^\mathrm{cl}[\varphi, {\tilde \varphi}]$ the field $\varphi \to
\phi + \varphi$, and collecting all cubic and quartic terms in the
fluctuating fields, i.e.\
\begin{align}
  S_{\rm int}^\mathrm{cl}[\varphi,{\tilde \varphi};\phi] =&
  -\frac{g}{2}\int_x {\tilde \varphi}_a(x)\varphi_a(x)\varphi_b(x)\varphi_b(x)
  \nonumber\\
  &  -\  g \int_x {\tilde \varphi}_a(x) \varphi_a(x)
  \varphi_b(x)\phi_b(x)
  \nonumber\\
  & -\ \frac{g}{2}\int_x {\tilde \varphi}_a(x)\phi_a(x)\varphi_b(x)\varphi_b(x).
  \label{eq:SqPhiXiclphi}
\end{align}
This can be compared to the respective expression for the quantum
theory, Eq.~(\ref{eq:SqPhiXiVev}).

\subsubsection{Correlation functions}
\label{sect:classcorr}
Comparing quantum and classical statistical dynamics amounts to
comparing the time evolution of classical statistical $n$-point
functions with that of the respective quantum ones. The classical
functions are obtained as phase-space averages
(\ref{eq:classaverage}) over trajectories given by solutions of
the classical field equation (\ref{eq:classicaleom}). As an
example, the macroscopic or average classical field is given by
\begin{equation}
  \phi_a^\mathrm{cl} (x)
  = \langle \varphi_a(x)\rangle_\mathrm{cl}
  = \int[{\rm d}\varphi_0^\mathrm{cl}] [\mathrm{d}\pi_0^\mathrm{cl}]  
     W[\varphi_0^\mathrm{cl},\pi_0^\mathrm{cl}]
    \varphi_a^\mathrm{cl}(x).
  \label{eq:phicl}
\end{equation}
This is equivalent to the field average obtained from the
generating functional (\ref{eq:ZClassPhiXi}) by functional
differentiation with respect to $\tilde J_a(x)$. Similarly, also
$\tilde{\phi}_a^\mathrm{cl}(x)$ can be obtained by functional
differentiation with respect to $J_a(x)$ in the same way as
described in Sect.~\ref{sec:QCorrFcts} if the quantum generating
functional in Eq.~(\ref{eq:WGenF}) is replaced by its classical
counterpart defined in Eq.~(\ref{eq:ZClassPhiXi}). Moreover, as
for the quantum case, $\tilde\phi_a(x)
\equiv 0$ for vanishing sources. This can be directly
seen by taking the functional derivative of $Z^\mathrm{cl}$ with
respect to $J_a(x)$, setting all sources to zero except $J_a(x)$,
and performing all steps backward in Eqs.~(\ref{eq:ZClassPhiXi})
and (\ref{eq:functionalasPI}). This is possible as the remaining
source term is linear in $\tilde\varphi$. Using the same reasoning
one finds that also the two-point correlation function $\tilde
F_{ab}(x,y)$, which is defined as in Eq.~(\ref{eq:propagators}) as
a functional derivative of $W^\mathrm{cl}=-i\ln Z^\mathrm{cl}$
with respect to $K^F$, vanishes identically in the absence of
sources.

Similarly, defining the connected classical statistical propagator
$F_{ab}^\mathrm{cl} (x,y)$ according to Eq.~(\ref{eq:propagators})
results in
\begin{align}
  &F_{ab}^\mathrm{cl} (x,y) + \phi_a^\mathrm{cl} (x) \phi_b^\mathrm{cl} (y)
  = \langle \varphi_a(x)\varphi_b(y)\rangle_\mathrm{cl}
  \nonumber\\
  &\quad= \int [{\rm d}\varphi_0^\mathrm{cl}][\mathrm{d}\pi_0^\mathrm{cl}]  
     W[\varphi_0^\mathrm{cl},\pi_0^\mathrm{cl}]
     \,\varphi_a^\mathrm{cl}(x)\varphi_b^\mathrm{cl}(y).
  \label{eq:Fcl}
\end{align}
The classical equivalent of the quantum spectral function
$\rho_{ab}^\mathrm{cl}(x,y)$ is obtained as follows: Taking the
functional derivatives according to Eq.~(\ref{eq:propagators})
leads, with Eq.~(\ref{eq:rhoinGRA}), to
\begin{align}
  &\rho_{ab}(x,y)
  = \int [{\rm d}\varphi_0^\mathrm{cl}][\mathrm{d}\pi_0^\mathrm{cl}]  
     W[\varphi_0^\mathrm{cl},\pi_0^\mathrm{cl}]
  \nonumber\\
  &\quad\times
     i\int\limits_{{\varphi_0=\varphi_0^\mathrm{cl}, }
                           {\pi_0=\pi_0^\mathrm{cl}}}
     \!\!\!\!\!\!
     \!{\cal D}'\varphi{\cal D}{\tilde\varphi}\,
     \big[\varphi_a(x)\tilde\varphi_b(y)-\tilde\varphi_a(x)\varphi_b(y)\big]
  \nonumber\\
  &\quad\times
     \,\exp\Bigg\{i\Bigg[S^\mathrm{cl}[\varphi,{\tilde \varphi}]
     +\int_{\mathbf{x}}\pi_{0,a}\tilde\varphi_{0,a}
  \Bigg]\Bigg\} {\cal J}[\varphi].
\label{eq:rhoclfromZcl}
\end{align}
If $x_0<y_0$ the expectation value of
$\varphi_a(x)\tilde\varphi_b(y)$ vanishes following the same
reasoning as for the field $\tilde\phi^\mathrm{cl}$, in accordance
with the fact that this term corresponds to the retarded
propagator $G^\mathrm{R}(x,y)$. Analogously, if $x_0>y_0$, the
advanced propagator, i.e. the average of
$\tilde\varphi_a(x)\varphi_b(y)$, vanishes.

We consider the case that $x_0=t_0$.
If $y_0>x_0$, one obtains from Eq.~(\ref{eq:rhoclfromZcl}) and the definition (\ref{eq:canmom}) of the canonical field momentum:
\begin{align}
  &\int [{\rm d}\varphi_0^\mathrm{cl}][\mathrm{d}\pi_0^\mathrm{cl}]  
     W[\varphi_0^\mathrm{cl},\pi_0^\mathrm{cl}]\,
     \frac{\delta}{\delta\pi^\mathrm{cl}_{0,a}(\mathbf{x})}
     \!\!\!\!\!
     \int\limits_{\quad {\varphi_0=\varphi_0^\mathrm{cl}, }
                           {\pi_0=\pi_0^\mathrm{cl}}}
     \!\!\!\!\!\!\!\!\!\!\!
     \!{\cal D}'\varphi{\cal D}{\tilde\varphi}\,\,
     \varphi_b(y)
  \nonumber\\
  &\quad\times
     \,\exp\Bigg\{i\Bigg[S^\mathrm{cl}[\varphi,{\tilde \varphi}]
     +\int_{\mathrm{x}}\pi_{0,a}\tilde\varphi_{0,a}
  \Bigg]\Bigg\} {\cal J}[\varphi]
  \nonumber\\
  &\ = -\int [{\rm d}\varphi_0^\mathrm{cl}][\mathrm{d}\pi_0^\mathrm{cl}]  
     W[\varphi_0^\mathrm{cl},\pi_0^\mathrm{cl}]
     \frac{\delta\varphi^\mathrm{cl}_b(y)}
          {\delta\pi^\mathrm{cl}_{0,a}(\mathbf{x})}
  \nonumber\\
  &\ = -\int [{\rm d}\varphi_0^\mathrm{cl}][\mathrm{d}\pi_0^\mathrm{cl}]  
     W[\varphi_0^\mathrm{cl},\pi_0^\mathrm{cl}]\,\int_\mathbf{z}
     \frac{\delta\varphi^\mathrm{cl}_a(t_0,\mathbf{x})}
          {\delta\varphi^\mathrm{cl}_{0,c}(\mathbf{z})}
     \frac{\delta\varphi^\mathrm{cl}_b(y)}
          {\delta\pi^\mathrm{cl}_{0,c}(\mathbf{z})} .
\label{eq:GAcl}
\end{align}
Extending this procedure to $t_0=y_0\le x_0$ one recovers the well-known fact that the classical spectral function is obtained by replacing $-i$
times the commutator with the Poisson bracket:
\begin{equation}
  \rho_{ab}^\mathrm{cl}(x,y)
  = -\langle\{\varphi_a(x),\varphi_b(y)\}_{\rm PB} \rangle_\mathrm{cl}.
  \label{eq:classspec}
\end{equation}
The Poisson bracket with respect to the initial fields is
\begin{eqnarray}
  \{\varphi_a(x),\varphi_b(y)\}_{\rm PB}
  &=& \int_\mathbf{z} \left[
  \frac{\delta \varphi_a(x)}{\delta\varphi_{0,c}(\mathbf{z})}
  \frac{\delta \varphi_b(y)}{\delta\pi_{0,c}(\mathbf{z})} \right.
  \nonumber\\
  && \quad\!\!\left. -\ \frac{\delta \varphi_a(x)}{\delta\pi_{0,c}(\mathbf{z})}
  \frac{\delta \varphi_b(y)}{\delta\varphi_{0,c}(\mathbf{z})}\right]
\end{eqnarray}
(Summation over $c$). Note that, in order to arrive at
Eq.~(\ref{eq:classspec}), we used that the Poisson brackets are
invariant under the classical time evolution of the fields and
therefore valid for any times $x_0$, $y_0$.

As a consequence, one finds the equal-time relations for the
classical spectral function:
$\rho_{ab}^\mathrm{cl}(x,y)|_{x_0=y_0} =
-i\sigma^2_{ab}\delta({\bf x}-{\bf y})$. Note that they are in
complete correspondence with the respective quantum relations in
Eq.~(\ref{eq:equaltimerho}). Equivalently, the free spectral
function $\rho_{0,cb}^\mathrm{cl}(x,y)$ and statistical function
$F_{0,cb}^\mathrm{cl}(x,y)$ are solutions of the homogeneous
equations corresponding to Eq.~(\ref{eq:F0rho0}), with initial
conditions determined for $\rho_0^\mathrm{cl}$ by the equal-time
canonical relations, and for $F_0^\mathrm{cl}$ by the initial
probability functional $W[\varphi_0,\pi_0]$. Also for the classical
statistical theory Eqs.~(\ref{eq:EOMphi}) and (\ref{eq:EOMFrho})
are the exact equations for the field $\phi$ and the correlation
functions $F$ and $\rho$, respectively. There is a difference
between the classical and quantum equations of motion only in the
self-energy contributions corresponding to $\Sigma^F$ and
$\Sigma^\rho$. This difference arises from the different
properties of the interaction part of the quantum and classical
actions (\ref{eq:SqPhiXi}) and (\ref{eq:SqPhiXicl}), respectively.
We discuss this difference for the non-perturbative 2PI $1/\cal N$
expansion to next-to-leading order in the following subsection.

Summarizing, one finds when comparing classical statistical and
quantum many-body dynamics that the generating functionals for
correlation functions are very similar in the classical and the
quantum theory. However, the quantum theory is characterized by
more vertices. As a consequence, the same techniques can be used
to derive time evolution equations of classical correlation
functions that are employed in quantum field theory. Eventually,
in the basis corresponding to the fields $\varphi$ and
$\tilde\varphi$, the classical dynamic equations have the same
form as their quantum analogues but are lacking certain terms due
to the reduced number of vertices.

\subsection{Quantum versus classical statistical evolution}
\label{sec:SelfEnClvsQ}

The classical statistical generating functional
(\ref{eq:ZClassPhiXi}) exhibits an important reparametrization
property: If the fluctuating fields are rescaled according to
\begin{align}
  \varphi_a(x) \to \varphi_a^\prime(x) 
  &=  \sqrt{g}\,\varphi_a(x),
  \nonumber\\
  \tilde{\varphi}_a(x) \to \tilde{\varphi}^\prime_a(x)
  &= (1/{\sqrt{g}})\,{\tilde{\varphi}_a(x)} \label{eq:rescaling}
\end{align}
then the coupling $g$ drops out of $S^\mathrm{cl}[\varphi,{\tilde
\varphi}]=S_0[\varphi,{\tilde \varphi}]+S_{\rm
int}^\mathrm{cl}[\varphi,{\tilde \varphi}]$ defined in
Eqs.~(\ref{eq:freeS}) and (\ref{eq:SqPhiXicl}). The free part
$S_0[\varphi,{\tilde \varphi}]$ remains unchanged and the
interaction part becomes
\begin{equation}
  S_{\rm int}^\mathrm{cl}[\varphi^\prime,{\tilde \varphi}^\prime]
  = -\frac{1}{2}\int_x {\tilde\varphi}^\prime_a(x)\varphi^\prime_a(x)\varphi^\prime_b(x)\varphi^\prime_b(x).
  \label{eq:SqPhiXiclRescaled}
\end{equation}
Moreover, the functional measure in Eq.~(\ref{eq:ZClassPhiXi}) is
invariant under the rescaling (\ref{eq:rescaling}), and the
sources can be redefined accordingly. Therefore, the classical
statistical generating functional becomes independent of $g$,
except for the coupling dependence entering the probability
distribution fixing the initial conditions. Accordingly, the
coupling does not enter the classical dynamic equations for
correlation functions. All the $g$-dependence enters the initial
conditions which are required to solve the dynamic equations.

In contrast to the classical case, this reparametrization property
is absent for the quantum theory: After the rescaling
(\ref{eq:rescaling}) one is left with
$S[\varphi^\prime,\tilde{\varphi}^\prime]$ whose coupling
dependence is given by the interaction part
\begin{eqnarray}
  S_{\rm int}[\varphi^\prime,\tilde{\varphi}^\prime]
  &=& -\frac{1}{2}\int_x {\tilde\varphi}^\prime_a(x)\varphi^\prime_a(x)\varphi^\prime_b(x)\varphi^\prime_b(x)
  \nonumber\\
  &&  -\frac{g^2}{8}\int_x {\tilde\varphi}^\prime_a(x){\tilde\varphi}^\prime_a(x)
                         {\tilde\varphi}^\prime_b(x)\varphi^\prime_b(x),
                         \quad
\label{eq:SqPhiXiRescaled}
\end{eqnarray}
according to Eq.~(\ref{eq:SqPhiXi}). Comparing to
(\ref{eq:SqPhiXiclRescaled}) one observes that the 'quantum'
vertex, which is absent in the classical statistical theory,
encodes all the $g$-dependence of the dynamics.

The comparison of quantum versus classical dynamics becomes
particularly transparent using the above rescaling. The rescaled
macroscopic field and statistical correlation function are given
by
\begin{equation}
\phi_a^\prime(x) = \sqrt{g} \phi_a(x) \,, \quad F_{ab}^\prime(x,y)
= g F_{ab}(x,y)\,, \label{eq:phiFRescaled}
\end{equation}
while the spectral function $\rho_{ab}(x,y)$ remains unchanged
according to Eqs.~(\ref{eq:propagators}) and (\ref{eq:rhoinGRA}).
Similarly, we define for the statistical self-energy entering the
dynamic equations (\ref{eq:EOMFrho})
\begin{equation}
\Sigma^{F\prime}_{ab}(x,y) = g \Sigma^F_{ab}(x,y)\,.
\label{eq:SigmaRescaled}
\end{equation}

\subsubsection{Quantum versus classical statistical self-energy}

\begin{figure}[tb]
\begin{center}
\includegraphics[width=0.45\textwidth]{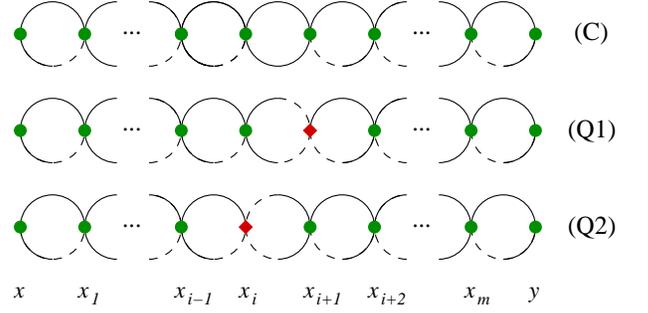}
\end{center}
\vspace*{-3ex} \caption{(color online) Diagrammatic representation
of the bubble chains contributing to the functions $I_{F}^\prime$
and $I_\rho$ at next-to-leading order in the 2PI $1/\cal N$
expansion. The meaning of lines and vertices is explained
in~Fig.~\ref{fig:DiagramsFGRGAHI}. The chains of type (Q1) and
(Q2) only appear for a quantum system. Type (C) is present both in
quantum and classical systems. These diagrams exhibit that in each
term contributing to the functions $I_F^\prime$ and $I_\rho$,
there is at most one loop involving two correlators $F'$ or
$\rho$. In the classical limit of quantum theory, the loops
involving two spectral functions $\rho$ are suppressed compared to
those with two statistical functions $F'$.}
\label{fig:DiagramsIBubbleChains}
\end{figure}

To identify the precise difference between the quantum and the
classical time evolution, details about the self-energies are
required. In the following we will employ the 2PI $1/{\cal N}$
expansion to next-to-leading order~\cite{Berges2002a,Aarts2002b}.
This is a nonperturbative expansion in powers of the inverse
number of field components $\cal N$ which, in the context of a
non-relativistic Bose gas, is ${\cal N}=2$ as discussed in detail
in Ref.~\cite{Gasenzer2005a}. We quote the result for the
self-energies for ${\cal N}=2$~\cite{Berges2002a,Aarts2002b},
which for the rescaled variables (\ref{eq:phiFRescaled}) and
(\ref{eq:SigmaRescaled}) read:
\begin{eqnarray}
\Sigma^{F\prime}_{ab}(x,y) & = & - \Big\{
 I_{F}^{\prime}(x,y) \phi^\prime_a (x)\phi^\prime_b(y)
 \nonumber\\
 && + \left[ I_{F}^{\prime}(x,y)+P_{F}^\prime(x,y)\right] F^\prime_{ab}(x,y)
  \nonumber\\
 && - \frac{g^2}{4} \left[ I_{\rho}(x,y)+P_{\rho}(x,y) \right] \rho_{ab} (x,y) \Big\},
 \nonumber\\[1.5ex]
 \Sigma^{\rho}_{ab} (x,y) & = &  - \Big\{
 I_{\rho}(x,y) \phi^\prime_a (x)\phi^\prime_b(y)
 \nonumber\\
 && + \left[ I_{\rho}(x,y)+P_{\rho}(x,y) \right] F^\prime_{ab}(x,y)
 \nonumber\\
 && + \left[ I_{F}^\prime(x,y)+P_{F}^\prime(x,y)\right] \rho_{ab} (x,y) \Big\}.
 \label{eq:SigmaNLO1N}
\end{eqnarray}
The functions $I_{F}^\prime$ and $I_{\rho}$ satisfy
\begin{eqnarray}
I_{F}^\prime(x,y) &=& - \Pi_F^\prime (x,y) \nonumber\\
&& + \int_{t_0}^{x_0} \! {\rm d}z\, I_{\rho}(x,z)\Pi_F^\prime
(z,y)
\nonumber\\
&& - \int_{t_0}^{y_0}\! {\rm d}z\, I_F^\prime(x,z) \Pi_{\rho}
(z,y),
 \nonumber\\[1.5ex]
 I_{\rho}(x,y) &=& - \Pi_{\rho} (x,y)
 \nonumber\\
 && + \int_{y_0}^{x_0}\! {\rm d}z\, I_{\rho}(x,z) \Pi_{\rho} (z,y),
 \label{eq:IFrho}
\end{eqnarray}
with
\begin{eqnarray}
 \Pi_F^\prime (x,y) & = & - \frac{1}{2}\Big[ F^\prime_{ab}(x,y) F^\prime_{ab}(x,y)
 \nonumber\\
 && - \frac{g^2}{4}\rho_{ab}(x,y) \rho_{ab}(x,y) \Big],
 \nonumber\\[1.5ex]
 \Pi_{\rho} (x,y) & = & - F^\prime_{ab} (x,y) \rho_{ab} (x,y).
 \label{eq:piself}
\end{eqnarray}
The functions $P_{F}^\prime$ and $P_\rho$, which vanish if $\phi =
0$, will be discussed below. Iterating the integral equation
(\ref{eq:IFrho}) implies that the functions $I_{F}^\prime$ and
$I_\rho$ can be represented as a series of bubble chains, as shown
in~Fig.~\ref{fig:DiagramsIBubbleChains}. A full line represents a
statistical correlator, while a retarded (advanced) propagator
line changes, from left to right, from full (broken) to broken
(full), see Fig.~\ref{fig:DiagramsFGRGAHI}.

The quantum and classical vertices are depicted analogously in
Fig.~\ref{fig:DiagramsFGRGAHI}(b). Connecting the correlators
through the respective vertices, one finds which types of bubble
chains appear. The classes of non-vanishing bubble chains shown in
Fig.~\ref{fig:DiagramsIBubbleChains} confirm the structure of
the functions $I_{F}^\prime(x,y)$ and $I_\rho(x,y)$ which are
determined by the integral equation (\ref{eq:IFrho}): In each term
contributing to the diagrammatic expansion of these functions
there is at most one loop containing two $F^\prime$ or two $\rho$
correlators, the latter resulting from either $G^\mathrm{R}$ or
$G^\mathrm{A}$. In addition to this one finds that only the
loop containing two $\rho$ correlators goes with the vertex which
is present in the quantum case only. Hence, considering the
classical dynamics, the $\sim g^2\rho^2$ terms are absent together
with the `quantum' vertex. For the same reasons, the contribution
$\sim (g^2/4) [I_\rho(x,y) + P_\rho(x,y) ]\rho_{ab}(x,y)$ to
$\Sigma^F_{ab}(x,y)$ is absent for the classical statistical
theory.

We proceed by considering the functions $P_F^\prime$ and $P_\rho$,
which are relevant for the case of a non-vanishing $\phi^\prime$.
The classes of bubble chains appearing in the self-energy for
$\phi^\prime\not = 0$ are derived from those shown in
Fig.~\ref{fig:DiagramsIBubbleChains}, by replacing at most one
full $F_{ab}^\prime(x,y)$ line by the product
$\phi_a^\prime(x)\phi_b^\prime(y)$ of mean fields. This means,
that the pair of correlators in one loop are replaced by either of
the combinations
\begin{eqnarray}
H_F^\prime(x,y) &=& -\phi_a^\prime(x) F_{ab}^\prime(x,y) \phi_b^\prime(y), \nonumber\\
H_\rho(x,y) &=& -\phi_a^\prime(x) \rho_{ab}(x,y) \phi_b^\prime(y).
\label{eq:HFHrho}
\end{eqnarray}
The functions $P_F^\prime$ and $P_\rho$ entering the self-energies
(\ref{eq:SigmaNLO1N}) then read~\cite{Aarts2002b}:
\begin{eqnarray}
 &&P_{F}^\prime (x,y) = - \Bigg\{ H_F^\prime (x,y)
 \nonumber\\
 &&\quad - \int_{t_0}^{x_0} {\rm d}z\, \left[ H_{\rho} (x,z) I_F^\prime (z,y) +
 I_{\rho} (x,z)H_F^\prime (z,y) \right]
 \nonumber\\
 &&\quad + \int_{t_0}^{y_0} {\rm d}z\, \left[  H_F^\prime (x,z) I_{\rho} (z,y) +
 I_F^\prime (x,z) H_{\rho} (z,y)\right]
 \nonumber\\
 &&\quad - \int_{t_0}^{x_0} {\rm d}z\, \int_{t_0}^{y_0} {\rm d}v\,
 I_{\rho} (x,z) H_F^\prime (z,v) I_{\rho} (v,y)
 \nonumber\\
 &&\quad + \int_{t_0}^{x_0} {\rm d}z\, \int_{t_0}^{z_0} {\rm d}v\,
 I_{\rho} (x,z) H_{\rho} (z,v) I_F^\prime (v,y)
 \nonumber\\
 &&\quad + \int_{t_0}^{y_0} {\rm d}z\, \int_{z_0}^{y_0} {\rm d}v\,
 I_F^\prime (x,z) H_{\rho} (z,v) I_{\rho} (v,y) \Bigg\},
\nonumber\\[1.5ex]
&& P_{\rho} (x,y) =  - g \Bigg\{ H_{\rho} (x,y)
 \nonumber\\
 &&\quad - \int_{y_0}^{x_0} {\rm d}z\,
 \left[ H_{\rho} (x,z) I_{\rho} (z,y) +
 I_{\rho} (x,z) H_{\rho} (z,y) \right]
 \nonumber\\
 &&\quad +  \int_{y_0}^{x_0} {\rm d}z\, \int_{y_0}^{z_0} {\rm d}v\,
 I_{\rho} (x,z) H_{\rho} (z,v) I_{\rho} (v,y) \Bigg\}.
\label{eq:Prho}
\end{eqnarray}
We observe that there are no differences in the functions
$P_F^\prime$ and $P_\rho$ in the classical statistical limit
except for a dependence on modified $I_F^\prime$ and $I_\rho$,
since the terms involving the mean field $\phi_a^\prime$
correspond either to $(F^\prime)^2$ or to $F^\prime\rho$ loops.
This implies that the presence of a non-vanishing mean field
$\phi^\prime$ does not add characteristic 'quantum' terms to the
dynamical equations of motion.

Moreover, the tadpole diagrams contain the classical vertex and
the $F^\prime$ correlator only. This shows that the left hand
sides of the quantum dynamic equations, Eqs.~(\ref{eq:EOMphi}) and
(\ref{eq:EOMFrho}), do not contain any contribution proportional
to the quantum vertex marked with a (red) square in
Fig.~\ref{fig:DiagramsFGRGAHI}(b). As a result, the dynamics in
the Hartree-Fock-Bogoliubov
\cite{Hartree1928a,Fock1930a,Bogoliubov1947a} approximation is the
same for the quantum and the classical statistical theory for same
initial conditions. Differences arise, at most, in the
self-energies $\Sigma^{F\prime}$ and $\Sigma^\rho$. We point out
that, neglecting the right-hand sides of Eqs.~(\ref{eq:EOMphi}),
(\ref{eq:EOMFrho}), these equations constitute a set of
time-dependent HFB equations for the mean field and the two-point
functions, cf., e.g.\ Refs.~\cite{Rey2004a,Gasenzer2005a}. In this
approximation, the dynamics of $\rho$ decouples from that of
$\phi$ and $F$. Neglecting also $F$, Eq.~(\ref{eq:EOMphi}) becomes
the Gross-Pitaevskii equation. Quantum fluctuations play no role
in these approximations. For this reason, HFB is commonly termed a
mean-field theory \cite{Fetter1971a}.

Summarizing, one concludes that all equations (\ref{eq:EOMFrho}), (\ref{eq:EOMphi}),
(\ref{eq:MofF}), (\ref{eq:SigmaNLO1N}), (\ref{eq:IFrho}), (\ref{eq:piself}),
(\ref{eq:HFHrho}), and (\ref{eq:Prho}) remain the same in the
classical statistical limit except for differing expressions for
the statistical components of the self-energy
\begin{eqnarray}
\Sigma^{F\prime}_{ab}(x,y) 
 & \stackrel{{\rm classical\,\,limit}}{\longrightarrow} 
 & - \Big\{
 I_F^\prime(x,y) \phi_a^\prime(x)\phi_b^\prime(y)
 \nonumber\\
 && + \left[ I_F^\prime(x,y)+P_F^\prime(x,y)\right] F_{ab}^\prime(x,y)\Big\},
 \nonumber\\
 \Pi_F^\prime(x,y) & \stackrel{{\rm classical\,\,
limit}}{\longrightarrow} & - \frac{1}{2} F_{ab}^\prime(x,y)
F_{ab}^\prime(x,y),
\label{eq:qmcl} \label{eq:SigmaNLO1NCl}
\end{eqnarray}
replacing the respective expression in Eqs.~(\ref{eq:SigmaNLO1N})
and (\ref{eq:piself}). The classical statistical self-energies can
be obtained from the respective quantum ones by dropping two
spectral ($\rho$-type) components compared to two statistical
($F$-type) functions. For vanishing macroscopic field $\phi$ where
$P_{F,\rho}=0$ this corresponds to the result of
Ref.~\cite{Aarts2002a}. Using the rescaled variables
(\ref{eq:phiFRescaled}) and (\ref{eq:SigmaRescaled}) the 'quantum'
terms can be directly identified since they are the only
$g$-dependent terms, which are absent in the classical statistical
theory according to the above discussion. As a consequence, for
the classical dynamics the effects of a larger coupling can always
be compensated by changing the initial conditions such that $F g$,
as well as $\phi \sqrt{g}$, remain constant. This cannot be
achieved once quantum corrections are taken into account, since
they become of increasing importance with growing coupling or
reduced initial values for $F$ and $\phi$.

\subsubsection{'Classicality' condition}

Eq.~(\ref{eq:SigmaNLO1NCl}) describes the differences between
quantum and classical statistical equations of motion. In turn one
can ask under which conditions these differences are negligible.
In that case the quantum dynamics can be well approximated by
classical statistical dynamics. To analyse this we iteratively
expand Eq.~(\ref{eq:IFrho}) in terms of $\Pi_F^\prime$ and
$\Pi_\rho$ and compare term by term the statistical components of
the quantum self-energies in Eqs.~(\ref{eq:SigmaNLO1N}) and
(\ref{eq:piself}) to the respective classical ones according to
(\ref{eq:SigmaNLO1NCl}). One finds that a sufficient condition for
the suppression of quantum fluctuations compared to classical
statistical fluctuations is given by
\begin{align}
  \left|F_{ab}^\prime(x,y)F_{cd}^\prime(z,w)\right|
  \gg\frac{3}{4} g^2 \left|\rho_{ab}(x,y)\rho_{cd}(z,w)\right| .
\label{eq:ClassCondNonEq}
\end{align}
This condition is not based on thermal equilibrium assumptions and
holds also for far-from-equilibrium dynamics. In particular, it is
independent of the value of the macroscopic field $\phi$. Our
results, therefore, agree with previous $\phi=0$ estimates in
Refs.~\cite{Aarts2002a,Berges2005a}. We also note that the
condition (\ref{eq:ClassCondNonEq}) is precisely the same as the
one obtained from the 2PI loop expansion
\cite{Berges2002a,Berges2005a}.

The condition (\ref{eq:ClassCondNonEq}) can be applied, of course,
also in thermal equilibrium, for which the statistical and
spectral correlation functions are no longer independent
quantities. In thermal equilibrium they are rather related to each
other through the fluctuation-dissipation relation, which, for a
homogeneous system in energy-momentum space, reads
\begin{align}
\label{eq:FlucDiss}
  F^\mathrm{(eq)\prime}(\omega,\mathbf{p})
  &= -i g \left(\frac{1}{2}+n(\omega,T)\right)\rho^\mathrm{(eq)}(\omega,\mathbf{p}),
\end{align}
with the Bose-Einstein distribution function
$n(\omega,T)=(e^{(\omega-\mu)/k_\mathrm{B}T}-1)^{-1}$. For large
temperatures, $k_\mathrm{B} T\gg\omega-\mu$, one has
$|F^{(\mathrm{eq})\prime}(\omega,\mathbf{p})|/g \gg
|\rho^\mathrm{(eq)}(\omega,\mathbf{p})|$, i.e., the classicality
condition is fulfilled for all modes whose occupation number $\sim
F^{(\mathrm{eq})\prime}(\omega,\mathbf{p})/g$ is much larger than
${\cal O}(1)$. The equivalent statement can be directly derived
from (\ref{eq:ClassCondNonEq}) also for nonequilibrium evolutions
whenever it is possible to define a suitable 'occupation number'
from a space-time or energy-momentum dependent proportionality
between $F$ and $\rho$. A nonequilibrium example will be discussed
in Sect.~\ref{sec:NumRes}.

Away from equilibrium the situation is often considerably more
complicated. Strictly speaking the condition
(\ref{eq:ClassCondNonEq}) must be valid at all times and for all
space points, or momenta in Fourier space, for the classical and
the quantum evolution to agree. In practice, however, it needs
only be fulfilled for time and space averages. In
Sect.~\ref{sec:NumRes} we will demonstrate how quantum evolution
can be approximated for not too late times by classical
statistical dynamics, if the correlation functions satisfy
(\ref{eq:ClassCondNonEq}) at initial time. In order to have
quantum fluctuations playing a significant role, also for
dynamically evolving gases, one either needs to increase the
interaction strength $g$ accordingly or change the phase-space
structure by changing the external trapping potential. For
example, in a one-dimensional trap, an effectively strong coupling
and strong quantum fluctuations can be induced by reducing the
line density of atoms while their interaction strength is kept
constant. Such a case will be considered in Sect.
\ref{sec:EquilCG} below.

\section{Far-from-equilibrium time evolution of an ultracold Bose gas}
\label{sec:NumRes}

In this section we apply the theoretical methods summarized above
to describe the equilibration dynamics of a uniform ultracold gas
of bosonic sodium atoms which are confined such that they can move
in one spatial dimension only. With present-day experimental
technology, such a situation is achievable, e.g., using strong
transversal confinement in an optical lattice or in a microtrap on
the surface of a chip.

We will compare the evolution involving only classical statistical
fluctuations with that which also takes into account quantum
corrections. Both, the classical and the quantum gas are assumed
to be initially characterized by the same initial conditions far
from thermal equilibrium. For the quantum gas, the ensuing
equilibration process is found to happen on two different time
scales. A fast dephasing period leads to a quasistationary state
which shows certain near-equilibrium characteristics but is still
far from being thermal. After this, the system approaches, within
an at least ten times longer period, the actual equilibrium state.
On the contrary, the classical gas does not show the dephasing
when considering the same initial mode occupation numbers as for
the quantum gas. We show that the dephasing in the short-time
evolution of the quantum gas can, to a certain extent, be
simulated by a classical gas if one chooses the same initial
values for the correlation functions. Our results show
explicitly that quantum fluctuations only play a role for modes
whose occupation number is sufficiently small. Hence, the dynamics
of the weakly interacting one-dimensional gas is almost purely
classical. The long-time evolution is different in the classical
and quantum cases, since only the latter can reach a Bose-Einstein
distribution. We then study the example of a strongly interacting
one-dimensional gas. For such a gas we find characteristic
differences. For instance, for given identical initial values for
the correlation functions, in the quantum evolution the decay of
correlations with the initial state takes much longer as it would
be expected from a calculation in the classical approximation.

\subsection{Initial conditions}
\label{sec:IniState}

The 2PI effective action approach is convenient for situations, where
at time $t=0$ one has a Gaussian state, i.e., a state, for which all
but the correlation functions of order one and two vanish \footnote{%
In the case that, at $t=0$, the $n$th-order connected correlation
function $\langle\Phi_{a_1}(0,{\bf x}_1)\cdots\Phi_{a_n}(0,{\bf x}
_n)\rangle_c$ is non-zero, with all $m$th-order functions, $m>n$,
vanishing, a straightforward generalization of the approach
involving the $n$PI effective action is at hand
\protect\cite{Berges2004a}. }. In the following we will consider a
one-dimensional uniform system, for which the two-point functions
$F_{ab}(x,y)$ and $\rho_{ab}(x,y)$ are spatially translation
invariant. We will therefore work in momentum space, where the
kinetic energy operator is diagonal. Moreover, we choose the mean
field $\phi$ to vanish initially. Then, for reasons of number
conservation, the equations of motion (\ref{eq:EOMphi}) and
(\ref{eq:EOMFrho}) will conserve $\phi=0$ for all times. We note
that, since there is no spontaneous symmetry breaking in one
spatial dimension at non-zero temperature, the field always
approaches zero eventually, irrespective of its initial value.

Having prescribed initial values $F_{ab}(0,0;p)$, with
$\rho_{ab}(0,0;p)$ given by Eq.~(\ref{eq:equaltimerho}), the
coupled system of integro-differential equations
(\ref{eq:EOMFrho}) yields the time evolution of the two-point
functions, in particular, of the momentum distribution
\begin{align}
\label{eq:npoft}
  n(t,p) = \frac{1}{2}\Big(F_{11}(t,t;p)+F_{22}(t,t;p)-\alpha\Big).
\end{align}
For the quantum gas, one has $\alpha=1$ from the Bose commutation
relations, while, for a gas following classical statistical
evolution, $\alpha=0$.

We choose, at $t=0$, a Gaussian momentum distribution
\begin{align}
\label{eq:npini}
  n(0,p) = \frac{n_1}{\sqrt{\pi}\sigma}e^{-p^2/\sigma^2}.
\end{align}
which constitutes a far-from-equilibrium state for the quantum gas
as well as for the classical gas if the interactions are non-zero
and the corresponding interaction energy is much larger than the
kinetic energy.

The initial pair correlation function vanishes,
\begin{align}
\label{eq:FmIniCond}
   0
   &=\ \frac{1}{2}\Big(F_{11}(t,t;p)-F_{22}(t,t;p)\Big)
    +iF_{12}(t,t;p),
\end{align}
for $t=0$, in accordance with total atom number conservation at non-relativistic energies
\footnote{
At non-relativistic energies, the total number of atoms in a finite system is fixed such that correlation functions defined as expectation values of a field operator product with an unequal number of creation and annihilation operators vanish identically.
}.
Hence,
\begin{align}
\label{eq:IniValF}
&F_{11}(0,0;p)=F_{22}(0,0;p)=n(0,p)+\alpha/2,
\\
&F_{12}(0,0;p)=F_{21}(0,0;p)\equiv0.
\end{align}
As far as the spectral functions are concerned, in the quantum case, the Bose commutation relations, and, in the classical case, the Poisson brackets require, cf.~Eqs.~(\ref{eq:equaltimerho}), (\ref{eq:classspec}):
\begin{align}
\label{eq:IniValrho}
\rho_{11}(t,t;p)=\rho_{22}(t,t;p)\equiv0&,
\\
-\rho_{12}(t,t;p)=\rho_{21}(t,t;p)\equiv1&.
\end{align}

We have investigated the dynamic evolution of a 1D Bose gas of
sodium atoms in a box of length $L=N_sa_s$, with periodic boundary
conditions. We choose the numerical grid such that it corresponds
to a lattice of $N_s$ points in coordinate space with grid
constant $a_s$, and the momenta on the Fourier transformed grid
are $p_n=(2/a_s)\sin(n\pi/N_s)$. The results presented in the
following are obtained using $N_s=64$ modes on a spatial grid with
grid constant $a_s=1.33\,\mu$m. We first consider a line density
of the atoms in the box of $n_1=10^7$ atoms$/$m. In this case the
atoms are weakly interacting with each other, such that
$g_\mathrm{1D}=\hbar^2\gamma n_1/m$, with the dimensionless
parameter $\gamma=1.5\cdot10^{-3}$. The width of the initial
momentum distribution is chosen to be
$\sigma=1.3\cdot10^5\,$m$^{-1}$. In order to explore a strongly
interacting gas we then reduce the total number of atoms by a
factor of $100$, increasing the dimensionless interaction
parameter $\gamma$ by $10^4$. Here it is important that we employ
a nonperturbative approximation which is not based on weak
interactions.

\subsection{Equilibration of the quantum gas}
\label{sec:EquilQG}

To solve Eqs.~(\ref{eq:EOMFrho}), with the self-energies given by
Eqs.~(\ref{eq:SigmaNLO1N}), for the initial conditions given in
the previous section, we have implemented a parallelized
Runge-Kutta solver and used a cluster of $3\,$GHz dual processor
PCs with up to one node per momentum mode. The correlation
functions $F_{ab}(t,t^\prime;p)$ and $\rho_{ab}(t,t^\prime;p)$
were propagated, for fixed $t^\prime$, along $t$, using a
second-order Runge-Kutta algorithm. After each Runge-Kutta step,
the $I_{F,\rho}$ integrals were updated according to
Eqs.~(\ref{eq:IFrho}). The dynamic equations derived from the 2PI
effective action are, by construction, number and energy
conserving. While number conservation, by virtue of the O(2)
symmetry of each diagram, is given exactly, energy conservation
may be violated by the chosen discretization along the time axis.
Hence, in order to ensure optimal energy conservation numerically,
a fourth-order Runge-Kutta algorithm was employed for the
propagation of the correlation functions along the diagonal
$t=t^\prime$.

\begin{figure}[tb]
\begin{center}
\includegraphics[width=0.45\textwidth]{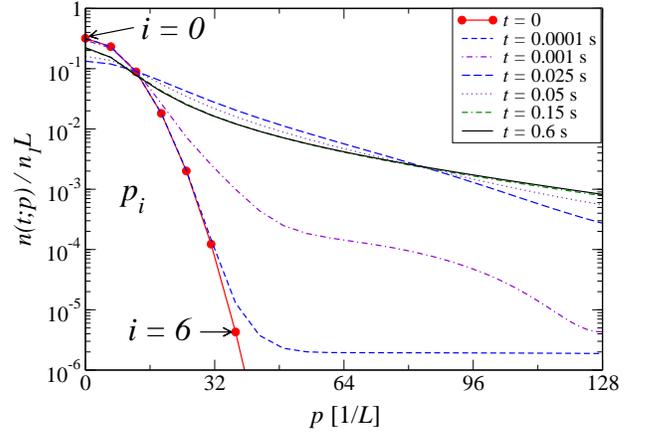}
\end{center}
\vspace*{-3ex} \caption{ (Color online) Momentum-mode distribution
$n(t;p)$ for the initial state (red filled circles, interpolated
by red solid line) and 6 subsequent times $t$ until no change can
be observed for $t>0.6\,$s. The interpolation of the final
distribution is shown as a black solid curve. Note the logarithmic
scale. The occupation number are normalized by the total number of
atoms in the box, $n_1L=853$. The gas is in a far-from-equilibrium
state initially, characterized by a Gaussian distribution
$n(0;p)$, Eq.~(\protect\ref{eq:npini}), with width
$\sigma=1.3\cdot10^5$m$^{-1}$. It is weakly interacting,
$\gamma=1.5\cdot10^{-3}$. Since we consider a homogeneous gas and
a symmetric initial state, the occupation numbers are invariant
under $p\to-p$. } \label{fig:n1ofp}
\end{figure}
\begin{figure}[tb]
\begin{center}
\includegraphics[width=0.45\textwidth]{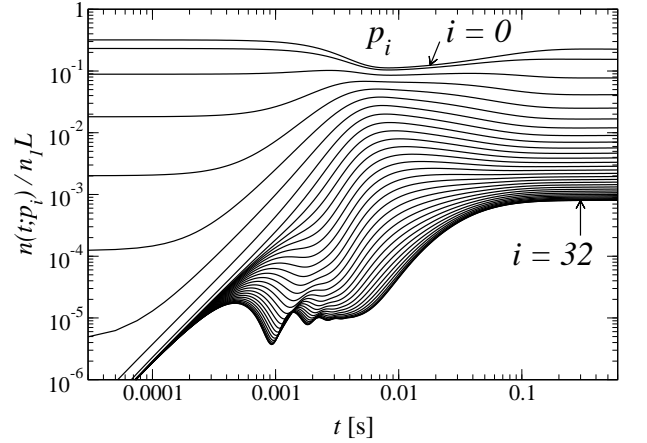}
\end{center}
\vspace*{-3ex}
\caption{
The normalized momentum-mode occupation numbers $n(t;p)/n_1L$,
corresponding to those shown in Fig.~\protect\ref{fig:n1ofp}, as
functions of time. Shown are the populations of the modes with
$p=p_i=2N_s/L\sin(i\pi/N_s)$, $i=0,1,...,N_s/2$, and one has
$n(t;-p)=n(t;p)$. A fast short-time dephasing period is followed
by a long quasistationary drift to the final equilibrium
distribution. Notice the double-logarithmic scale. }
\label{fig:n1oft}
\end{figure}
Fig.~\ref{fig:n1ofp} shows, as a (red) solid curve, the initial
Gaussian momentum distribution of the gas, on a logarithmic scale,
where it forms an inverted parabola. The filled circles indicate
the numerically calculated modes $p_i$. In the same figure, the
time evolution of the distribution is shown for different times
between $t=0.1\,$ms and $0.6\,$s. For times greater than about
$0.15\,$s, there is only very little change observed. As a
function of time, the evolution of the single mode occupations is
shown in Fig.~\ref{fig:n1oft}. We observe that the system very
quickly, after about $5\,\mu$s, evolves to a quasistationary
state, and that the subsequent drift to the equilibrium
distribution takes roughly ten times longer. In passing we note
that the mean-field Hartree-Fock (HF) approximation, for which
$\Sigma^F=\Sigma^\rho\equiv0$ in Eqs.~(\ref{eq:EOMFrho}),
conserves exactly all mode occupations and no equilibration is
seen.

\begin{figure}[tb]
\begin{center}
\includegraphics[width=0.45\textwidth]{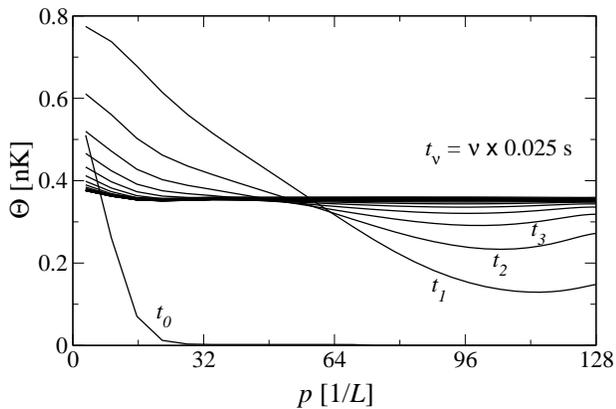}
\end{center}
\vspace*{-3ex} \caption{Momentum and time dependent temperature variable $\Theta(t;p)$
obtained by fitting the distribution
$n(t;p)=[\exp\{(\omega(p)-\mu)/k_B\Theta(t;p)\}-1]^{-1}$ to the
distribution obtained from the results shown in
Fig.~\protect\ref{fig:n1oft}, for different, equally spaced times
between $t=0$ and $t=0.6\,$s. One observes that, during the
quasistationary period, $0.01\,$s$<t<0.1\,$s, no temperature can
be associated to the distribution. Only at very large times,
$\Theta$ becomes approximately $p$-independent.}
\label{fig:betaofp}
\end{figure}

In order to estimate to which extent the final distribution
approaches that of the actual equilibrium state of the gas, we
fitted the distribution to the Bose-Einstein-like form
$n(t;p)=[\exp\{(\omega(p)-\mu)/k_B\Theta(t;p)\}-1]^{-1}$, with a
$p$-dependent temperature variable $\Theta(t;p)$. Here $\omega(p)$
was derived from the time-derivatives of the statistical function
$F(t,t';p)$ at $t=t'$. If a Bose-Einstein distribution is
approached the temperature can be obtained from the slope of
$\log(n^{-1} + 1)$ and the chemical potential $\mu$ from its value
at $\omega=0$. Fig. \ref{fig:betaofp} shows $\Theta(t;p)$ for
$t=0...0.6\,$s. Obviously, during the quasistationary drift
period, no temperature can be attributed to $n(t;p)$, while, for
large $t$, $\Theta$ becomes approximately $p$-independent
\footnote{A rigorous calculation of the final temperature can be
based on the Bose-Einstein distribution defined in energy-momentum
space, which may be extracted using the fluctuation-dissipation
(\ref{eq:FlucDiss}) at late times along the lines of
Ref.~\cite{Berges2003a}.}. We deduce an approximate final
temperature from $\Theta(0.6s;128/L)=T=0.35$nK with
$\mu=1.08\,g_\mathrm{1D}n_1$ for the above given values of
$g_\mathrm{1D}$ and $n_1$, which, hence, deviates from the HFB result $\mu=g_\mathrm{1D}n_1$ by 8\% only.

We furthermore studied, in the spirit of Refs.~\cite{Berges2001a,Aarts2001a}, the time dependence of the ratio of the envelopes of the unequal-time correlation functions, specifically, $\xi(t;p)=[(F_{11}(t,0;p)^2+F_{12}(t,0;p)^2)/(\rho_{11}(t,0;p)^2+\rho_{12}(t,0;p)^2)]^{1/2}/n(t;p)$.
\begin{figure}[tb]
\begin{center}
\includegraphics[width=0.45\textwidth]{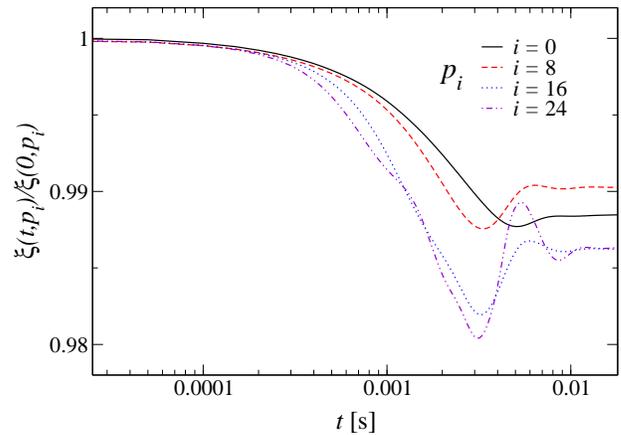}
\end{center}
\vspace*{-3ex}
\caption{
(Color online)
Ratio of the envelopes of the unequal-time correlation functions, $\xi(t;p)=[(F_{11}(t,0;p)^2+F_{12}(t,0;p)^2)/(\rho_{11}(t,0;p)^2+\rho_{12}(t,0;p)^2)]^{1/2}/n(t;p)$, for four different momentum modes, as a function of time.
Due to the normalization with respect to $n(t;p)$ all $\xi(t;p)$ are of the same order of magnitude.
$\xi$ is a measure of the interdependence of the statistical and spectral functions, and its settling to a constant value during the quasistationary drift period indicates that these functions become, as in thermal equilibrium, connected through a fluctuation-dissipation relation.
}
\label{fig:flucdiss}
\end{figure}
Fig. \ref{fig:flucdiss} shows $\xi$, for four different momentum modes, as a function of time.
Due to the normalization with respect to $n(t;p)$ all $\xi(t;p)$ are of the same order of magnitude.
However, they show a distinct time evolution during the dephasing period, before they settle to a constant value during the quasistationary drift.
$\xi$ is a measure of the interdependence of the statistical and spectral functions, which, in thermal equilibrium, are connected through the fluctuation-dissipation relation \cite{Aarts2001a}.
In the momentum-frequency domain, this relation
is given in Eq.~(\ref{eq:FlucDiss}).
Hence, the stationary $\xi$ indicates that $F$ and $\rho$ are linked to each other long before the momentum distribution becomes thermal.
\begin{figure}[tb]
\begin{center}
\includegraphics[width=0.45\textwidth]{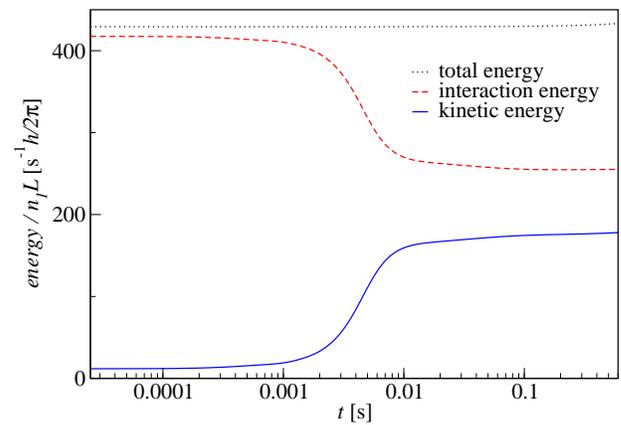}
\end{center}
\vspace*{-3ex}
\caption{
(Color online)
Evolution of the kinetic and interaction contributions to the total energy of the gas.
During the quasistationary drift these contributions assume the same order of magnitude, calling in mind the virial theorem.}
\label{fig:energies}
\end{figure}

A further signature is found when comparing the kinetic and interaction contributions to the total energy as shown in Fig.~\ref{fig:energies}.
During the quasistationary drift, these contributions are constant and of the same order of magnitude, calling in mind the virial theorem.

In summary, during the drift period, the system is not yet in
equilibrium as far as the momentum distribution and temperature is
concerned, but shows important characteristics of a system close
to equilibrium.

\subsection{Evolution of the classical gas}
\label{sec:EquilCG}

In Section \ref{sec:ClvsQ} we have discussed in detail the
distinction between the quantum and classical statistical
contributions to the 2PI effective action and to the dynamic
equations. In the following we compare the predictions for the
classical statistical theory with those presented in the preceding
section, and pointing to the distinct differences.

We solved Eqs.~(\ref{eq:EOMFrho}), with the self-energies now
given by Eqs.~(\ref{eq:SigmaNLO1NCl}), for the initial conditions
given in Section \ref{sec:IniState}, first with $\alpha=0$. For
comparison we also consider the case $\alpha=1$ such that the
classical and quantum initial correlation functions $F$ and $\rho$ are identical.
In the classical case, the self-energies as well as the equations
(\ref{eq:IFrho}) determining the coupling functions $I_{F,\rho}$
are lacking certain terms compared to their quantum counterparts,
cf.~Eq.~(\ref{eq:SigmaNLO1NCl}).

\begin{figure}[tb]
\begin{center}
\includegraphics[width=0.45\textwidth]{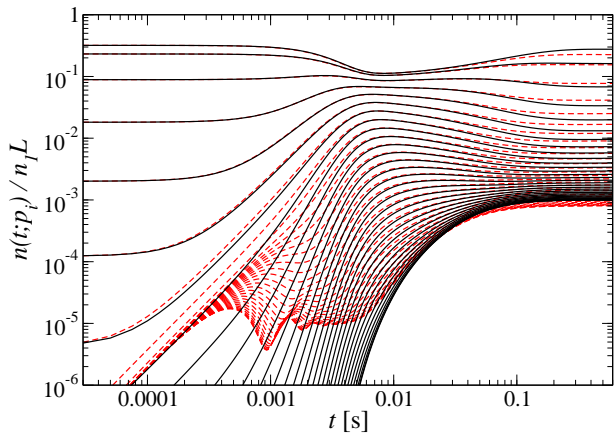}
\end{center}
\vspace*{-3ex} \caption{ (Color online) The normalized
momentum-mode occupation numbers $n(t;p)/n_1L$ for the classical
gas (black solid lines) compared to their quantum counterparts
from Fig.~\protect\ref{fig:n1oft} (red dashed lines), as functions
of time. Shown are the populations of the modes with
$p=p_i=2N_s/L\sin(i\pi/N_s)$, $i=0,1,...,N_s/2$, and one has
$n(t;-p)=n(t;p)$. In contrast to the quantum statistical evolution
there is no quick dephasing in the classical case, such that the
initially empty modes become only subsequently filled, as is
discussed in the main text. At large times, the classical and
quantum gases necessarily evolve to different distributions. }
\label{fig:n1oftCl}
\end{figure}

The time evolution of the initially Gaussian far-from-equilibrium
momentum distribution of the classical gas is shown in
Fig.~\ref{fig:n1oftCl}. The mode occupations are shown as (black)
solid lines, and for better comparison, we have quoted the quantum
evolution from Fig.~\ref{fig:n1oft} as (red) dashed curves.
One observes that the time evolution of the modes with occupation
number $n(t;p)>1$, i.e., $n(t;p)/n_1L>10^{-3}$, is, for most of
the time, identical to that obtained in the quantum case,
confirming condition (\ref{eq:ClassCondNonEq}). As expected, for
the strongly populated modes of a weakly interacting bosonic gas
quantum fluctuations do not play a significant role for not too
large times. Only when the evolution approaches the equilibrium
state, the differences between quantum and classical statistics
are expected to lead to a Bose-Einstein and classical
distribution, respectively. We point out that, although we have
chosen, for our comparison, the same initial occupation numbers in
the two cases, the total energies are different since the initial
correlation functions differ according to Eq.~(\ref{eq:IniValF}).
Hence, also the final-state occupation numbers of the low momentum
modes can differ.

We finally point to the substantial differences in the short-time
evolution of the weakly populated modes. While, in the quantum
gas, the large-momentum mode populations are all growing at the
same rate, the modes of the classical gas become populated much
more gradually. The quantum-gas modes are occupied by $0.01$ and
$1$ particle per mode already between $0.5$ and $1\,\mu$s, while
the classical modes need up to ten times longer. The distinct
quantum behaviour of the modes can be understood as follows. At
the energies present in such an ultracold gas, the atomic
interactions are essentially pointlike, i.e., the range of the
potential is not resolved and the coupling function or scattering
amplitude is constant over the range of relevant momenta. Hence,
in a single scattering event, the distance of two atoms is
localized to zero, such that the relative momentum of the atoms is
completely unknown immediately after the collision. This means
that the transfer probability of the atoms is the same for any
final momentum mode, and this is observed as a quick, collective
population in the respective, so far essentially unoccupied modes.

For comparison we repeated our calculations for a non-local
interaction potential. In the momentum domain, this corresponds to
a coupling function which is cut off at large momenta, and we
chose the cutoff within the range of the momenta shown explicitly
in Fig.~\ref{fig:n1oftCl}. In this case we find that the quantum
evolution is modified such that it becomes similar to that of the
classical gas. In particular, all modes above the cutoff populate
gradually one after each other.

\begin{figure}[tb]
\begin{center}
\includegraphics[width=0.45\textwidth]{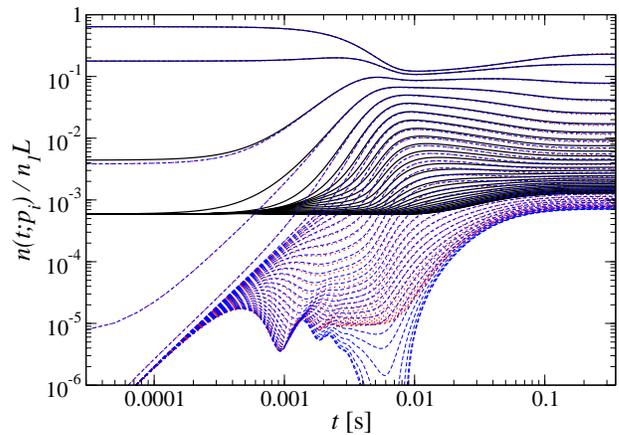}
\end{center}
\vspace*{-3ex}
\caption{
(Color online)
The normalized momentum-mode occupation numbers $n(t;p)/n_1L$ for the classical gas (black solid lines) compared to their quantum counterparts from Fig.~\protect\ref{fig:n1oft} (red dotted lines), as functions of time.
All parameters are chosen as in Fig.~\protect\ref{fig:n1oftCl}, except for $\sigma=6.5\cdot10^4\,$m$^{-1}$.
The dashed (blue) lines show $n(t,p)-1/2$, i.e., after subtracting the additional flat initial distribution which simulates the quantum ``zero point fluctuations''.
}
\label{fig:n1oftClqinic}
\end{figure}
The differences between the quantum and classical evolutions shown
above depend, however, considerably on the choice of initial
conditions. To compare the characteristics of the evolutions which
are independent of the initial choice of $F$, we have repeated the
classical calculations for an initial momentum distribution where,
as compared to before, a constant occupation number $1/2$ has been
added. Hence, we chose $\alpha=1$ in the initial values of $F$,
Eq.~(\ref{eq:IniValF}), as in the quantum case, such that $F$ is
identical for $x_0=y_0=t_0$ in the classical and quantum cases.
The results are shown in Fig.~\ref{fig:n1oftClqinic}. The (red)
dotted lines show, again, the quantum evolution, while the
classical mode populations for the same initial conditions for $F$
and $\rho$ are shown as solid (black) lines. Subtracting $1/2$
from each $F_{aa}(t,t;p)$ gives the dashed (blue) lines. We find,
that during the initial period the evolution of the variation of
the high-momentum modes with respect to their initial occupation
is identical to the quantum evolution of the occupation numbers.
At intermediate times, however, there are deviations which lead to
occupation numbers up to a percent lower than $1/2$. Although the
chosen initial conditions which, in the classical case correspond
to a base occupation of each mode with $1/2$ atom, seem
unphysical, our results show that for the dilute, weakly
interacting gas under consideration, there are differences only in
those modes which, in the mean, are populated with less than one
atom. Quantum statistical fluctuations play a role only for these
modes.

\begin{figure}[tb]
\begin{center}
\includegraphics[width=0.45\textwidth]{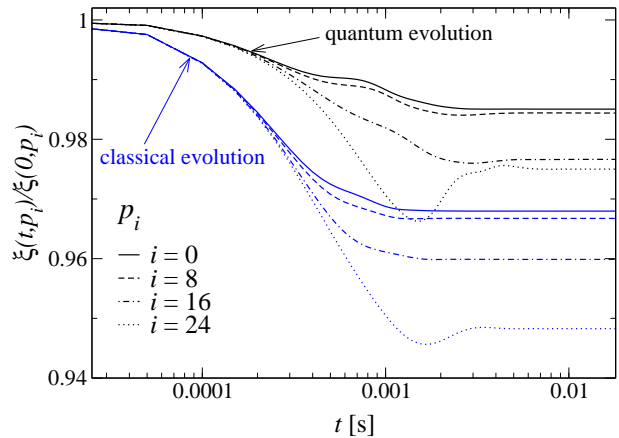}
\end{center}
\vspace*{-3ex} \caption{ (Color online) Ratio of the envelopes of
the unequal-time correlation functions,
$\xi(t;p)=[(F_{11}(t,0;p)^2+F_{12}(t,0;p)^2)/(\rho_{11}(t,0;p)^2+\rho_{12}(t,0;p)^2)]^{1/2}/n(t;p)$,
for four different momentum modes, as a function of time. This
figure is analogous to Fig.~\ref{fig:flucdiss}, but shows the
difference of the quantum (black lines) and classical (blue lines)
evolutions for a gas in the strongly correlated regime
$\gamma=15$, for $n_1=10^5\,$m$^{-1}$. The results show that in
this regime, where quantum and classical statistical evolution
differ considerably, the quantum evolution conserves information
about the initial conditions much longer, here roughly by an order
of magnitude in time. } \label{fig:flucdisstonks}
\end{figure}

The situation is very different for an ultracold gas in the
strongly interacting regime. We obtain this from the case
discussed so far by reducing the total number of atoms by a factor
of $100$ and increasing the dimensionless interaction parameter
$\gamma$ by $10^4$. In this case the 'classicality' condition
(\ref{eq:ClassCondNonEq}) is clearly violated and we expect strong
corrections due to quantum fluctuations. We emphasize that here
our nonperturbative approximation is crucial in order to be able
to consider such strongly interacting systems. Note that the
structure of the classical equations of motion,
Eqs.~(\ref{eq:EOMFrho}), with the self-energies given by
Eqs.~(\ref{eq:SigmaNLO1NCl}), is such that they are invariant
under a simultaneous rescaling of $\gamma$ and $n_1$ which leaves
$\gamma n_1^2$ unchanged. The quantum equations will, however
change, with the specific quantum terms becoming more and more
important with growing $\gamma$ and reduced density $n_1$. We
compare the evolution, again, for identical initial conditions as
in the case shown in Fig.~\ref{fig:n1oftClqinic}. The deviations
between quantum and classical statistical evolution are now
quantitatively substantial. While many qualitative aspects are
similar to those discussed above, now all momentum modes show
clear deviations. As a characteristic example, we present, in
Fig.~\ref{fig:flucdisstonks}, the ratio of the envelopes of the
unequal-time correlation functions, in analogy to
Fig.~\ref{fig:flucdiss}, for the quantum and classical evolution
of the strongly interacting gas. One observes that the
unequal-time correlation function decays more rapidly in the
classical statistical case. Our results illustrate that the
quantum system keeps much longer the information about the initial
conditions, here roughly by an order of magnitude in evolution
time. On the other hand, the example also shows that the classical
statistical system still behaves qualitatively similar to the
quantum gas, despite the low densities and strong interactions. We
would like to emphasize that clearly distinguishing quantum and
classical statistical fluctuations will typically require high
precision, both from the calculational as well as experimental
point of view.

\section{Conclusions}
\label{sec:Concl}

In this article we have studied the non-equilibrium dynamics of an
ultracold Bose gas and compared the theoretical predictions from a
full quantum approach with those where only classical statistical
fluctuations were taken into account. On the basis of functional
techniques in quantum field theory we reviewed the difference in
the generating functionals, the 2PI effective actions which
determine the dynamic equations, between the quantum and the
classical cases. The functional descriptions for both cases show
to be very similar. In particular, the crucial difference is the
absence, in the classical versus the quantum case, of certain
coupling terms or vertices in the action which defines the theory
and which enters the generating functional of correlation
functions. As a consequence, the classical generating functional
is characterized by an important reparametrization property, such
that for the classical dynamics the effects of a larger
self-interaction can always be compensated by a smaller density.
Quantum corrections violate this invariance property. They become
of increasing importance with growing scattering length or reduced
density. We have used this to derive a 'classicality' condition,
which is not based on thermal equilibrium assumptions and holds
also for far-from-equilibrium dynamics. In particular, it is
independent of the value of the macroscopic field $\phi$.

To illustrate the possibilities one has at hand, we studied the
equilibration dynamics of a homogeneous ultracold gas of
interacting sodium atoms in one spatial dimension. The gas is
assumed to be, initially, in a state far from thermal equilibrium,
characterized by a Gaussian momentum distribution centred at zero
momentum. Extending on earlier results \cite{Gasenzer2005a} we
find that the evolution takes place via two distinct periods, with
a fast initial dephasing followed by a slow quasistationary drift
to the final equilibrium distribution. While the gas, during the
quasistationary drift, already shows important characteristics of
a near-equilibrium situation like a fluctuation-dissipation
relation and constancy of the total kinetic energy, no temperature
can yet be associated to the momentum distribution.

We compared this evolution to the evolution of a classical gas
with different initial conditions. If the initial mode occupation
numbers are chosen as in the quantum case, no short-time dephasing
period is found. The modes are rather populated subsequently, with
the quasistationary drift setting in gradually. The long-time
approach to the final equilibrium state expectedly shows a
distinct behaviour compared to the quantum gas. We moreover showed
that the dephasing in the short-time evolution of the quantum gas
can be approximately simulated by a classical gas if one chooses
appropriate initial conditions. Our results demonstrate explicitly
that quantum fluctuations only play a role for modes whose
occupation number is sufficiently small. We found that the
dynamics of the weakly interacting one-dimensional gas is almost
purely classical. In contrast, for a strongly interacting
one-dimensional gas we observed substantial quantitative
deviations. There are distinctive properties of quantum versus
classical statistical dynamics, as e.g.\ that, given identical
initial values for the correlation functions, in the quantum
evolution, information about the initial conditions is conserved
much longer as it would be expected from a calculation in
classical approximation. Such differences may be expected to be of
interest in the context of strongly correlated ultracold atomic
gases, e.g. in lower-dimensional traps or near Feshbach scattering
resonances. However, we emphasize that high precision measurements
together with accurate calculations are typically required to be
able to clearly distinguish between effects of quantum and
classical statistical fluctuations. We think that the
nonperturbative methods presented here can be a very valuable tool
to precisely identify the effect of quantum fluctuations on the
time evolution of ultracold quantum gases observed in present-day
and near-future experiments.

\acknowledgments \noindent We thank Gert Aarts for collaboration
on related work, and Hrvoje Buljan, Thorsten K\"ohler, Markus
Oberthaler, Jan Martin Pawlowski, J\"org Schmiedmayer, Gora
Shlyapnikov, and Kristan Tem\-me for inspiring and valuable
discussions, and Werner Wetzel for his continuing support
concerning computers. This work has been supported by the Deutsche
For\-schungs\-gemeinschaft.

\begin{appendix}

\section{Initial conditions}
\label{app:inistate}
In this appendix we provide explicit expressions for a general Gaussian initial-state density matrix $\rho_D[\varphi_0+\tilde\varphi_0/2,\varphi_0-\tilde\varphi_0/2]$ and a Gaussian probability functional $W[\varphi_0,\pi_0]$ entering the generating functionals (\ref{eq:ZQuantPhiXi}) and (\ref{eq:ZClassPhiXi}), respectively. 
These specify the initial conditions for the dynamic equations  (\ref{eq:EOMFrho}), (\ref{eq:EOMphi}).
We provide expressions for a spatially homogeneous system, as considered in Sect.~\ref{sec:NumRes}.

The most general Gaussian initial density matrix takes, in the representation (\ref{eq:RTransf}) of the fields, the form
\begin{align}
  &\rho_D[\varphi_0+\tilde\varphi_0/2,\varphi_0-\tilde\varphi_0/2]
  = \frac{1}{2\pi\xi_1\xi_2}
  \exp\Big\{
  -\tilde\varphi_{0,a}\sigma^2_{ab}\phi_{0,b}
  \nonumber\\
  &\quad -\frac{1}{2\xi_a^2}(\varphi_{0,a}-\phi_{0,a})^2
     +i\frac{\eta_a}{\xi_a}(\varphi_{0,a}-\phi_{0,a})\tilde\varphi_{0,a}
  \nonumber\\
  &\quad -\frac{\sigma_a^2}{8\xi_a^2}\tilde\varphi_{0,a}^2\Big\},
\end{align}
where summation over $a$ and $b$ is implied in the exponent.
Since $\rho_D$ involves a factor of the above form above for each momentum mode,  mode indices at the fields and parameters, as well as summation over momenta in the exponent have been neglected.
In order to reflect the symmetry (\ref{eq:canmom}) between the field and the canonical momentum, the six parameters $\xi_a$, $\eta_a$, and $\sigma_a$ are reduced to three independent parameters through the conditions
\begin{align}
  \sigma_1
  &=\sigma_2\equiv\sigma,
  \nonumber\\
  \xi_1^2
  &= \eta_2^2+{\sigma^2}/{4\xi_2^2},
  \nonumber\\
  \eta_1
  &= -\eta_2{\xi_2}/{\xi_1}.
\end{align}
Using the definition of the initial statistical correlation functions in terms of the initial density matrix,
\begin{align}
  F_{ab}(0,0)
  &= \frac{1}{2}\mathrm{Tr}\Big[\rho_D(t_0)\{\Phi_a(0),\Phi_b(0)\}\Big]
     -\phi_{0,a}\phi_{0,b},
\end{align}
and inserting unit operators $\int[\mathrm{d}\varphi]|\varphi\rangle\langle\varphi|$ and/or $\int[\mathrm{d}\pi]|\pi\rangle\langle\pi|$ one finds that the three free parameters are related to the initial correlation functions as follows
\begin{align}
  \xi_1^2
  &= F_{11}(t_0,t_0)-\phi_{0,1}\phi_{0,1},
  \nonumber\\
  \xi_1\eta_1
  &= F_{12}(t_0,t_0)-\phi_{0,1}\phi_{0,2},
  \nonumber\\
  \eta_1^2+\sigma_1^2/4\xi_1^2
  &= F_{22}(t_0,t_0)-\phi_{0,2}\phi_{0,2}.
  \label{eq:Finxietasigma}
\end{align}
Again, all mode labels have been suppressed.
The initial spectral functions $\rho_{ab}(0,0)$ do not enter the density matrix as they are fixed, in the quantum and classical statistical cases, by the commutation relations and the Poisson brackets, respectively.

Note that the initial conditions chosen in Sect.~\ref{sec:IniState} for the numerical evaluation of the dynamics of a one-dimensional Bose gas, with $F_{11}(0,0)=F_{22}(0,0)$, $F_{12}(0,0)=0$ and $\phi_{0,a}\equiv0$ correspond to $\xi_1^2=\xi_2^2\equiv\xi^2=F_{11}(0,0)$, $\eta_a=0$, $\sigma=2\xi^2$, such that
\begin{align}
  &\rho_D[\varphi_0+\tilde\varphi_0/2,\varphi_0-\tilde\varphi_0/2]
  \nonumber\\
  &\quad = \frac{1}{2\pi\xi^2}
  \exp\left\{
  -\frac{1}{2\xi^2}\varphi_{0,a}^2
  -\frac{\xi^2}{2}\tilde\varphi_{0,a}^2\right\}.
\end{align}

We close this appendix by providing the expression for the probability functional $W[\varphi_0,\pi_0]$ obtained through the inverse of the Fourier transform (\ref{eq:FTrhoW}):
\begin{align}
  &W[\varphi_0,\pi_0]
  \nonumber\\
  &\  
  = \frac{1}{\pi^2\sigma^2}
  \exp\left\{-\frac{2\xi_a^2}{\sigma^2}
        \left(\frac{\eta_a}{\xi_a}\varphi_{0,a}-\pi_{0,a}\right)^2
	-\frac{1}{2\xi_a^2}\varphi_{0,a}^2\right\}.
\end{align}
(Summation over $a$ in the exponent).
It is straightforward to show that the above expression is symmetric under exchange of $\varphi_{0,a}$ and $\pi_{0,a}$, and that, by inserting $\pi_{0,a}=i\sigma^2_{ab}\varphi_{0,b}$, one obtains the expression
\begin{align}
  &W[\varphi_0]
  = \frac{1}{(2\pi)^2|F_{ab}|}
  \exp\Big\{-\frac{1}{|F_{ab}|}\Big[
  F_{11}(0,0)\,\varphi_{0,2}^2
  \nonumber\\
  &\qquad  
  +F_{22}(0,0)\,\varphi_{0,1}^2
  -2F_{12}(0,0)\,\varphi_{0,1}\varphi_{0,2}\Big]\Big\},
\end{align}
with $|F_{ab}|\equiv F_{11}(0,0)F_{22}(0,0)-F_{12}(0,0)F_{21}(0,0)$.

\end{appendix}

\bibliography{mybib,additions}

\begin{thebibliography}{73}
\expandafter\ifx\csname natexlab\endcsname\relax\def\natexlab#1{#1}\fi
\expandafter\ifx\csname bibnamefont\endcsname\relax
  \def\bibnamefont#1{#1}\fi
\expandafter\ifx\csname bibfnamefont\endcsname\relax
  \def\bibfnamefont#1{#1}\fi
\expandafter\ifx\csname citenamefont\endcsname\relax
  \def\citenamefont#1{#1}\fi
\expandafter\ifx\csname url\endcsname\relax
  \def\url#1{\texttt{#1}}\fi
\expandafter\ifx\csname urlprefix\endcsname\relax\def\urlprefix{URL }\fi
\providecommand{\bibinfo}[2]{#2}
\providecommand{\eprint}[2][]{\url{#2}}

\bibitem[{\citenamefont{Cornell and Wieman}(2002)}]{Cornell2002a}
\bibinfo{author}{\bibfnamefont{E.~A.} \bibnamefont{Cornell}} \bibnamefont{and}
  \bibinfo{author}{\bibfnamefont{C.~E.} \bibnamefont{Wieman}},
  \bibinfo{journal}{Rev. Mod. Phys.} \textbf{\bibinfo{volume}{74}},
  \bibinfo{pages}{875} (\bibinfo{year}{2002}).

\bibitem[{\citenamefont{Leggett}(2001)}]{Leggett2001a}
\bibinfo{author}{\bibfnamefont{A.~J.} \bibnamefont{Leggett}},
  \bibinfo{journal}{Rev. Mod. Phys.} \textbf{\bibinfo{volume}{73}},
  \bibinfo{pages}{307} (\bibinfo{year}{2001}).

\bibitem[{\citenamefont{Dalfovo et~al.}(1999)\citenamefont{Dalfovo, Giorgini,
  Pitaevskii, and Stringari}}]{Dalfovo1999a}
\bibinfo{author}{\bibfnamefont{F.}~\bibnamefont{Dalfovo}},
  \bibinfo{author}{\bibfnamefont{S.}~\bibnamefont{Giorgini}},
  \bibinfo{author}{\bibfnamefont{L.~P.} \bibnamefont{Pitaevskii}},
  \bibnamefont{and}
  \bibinfo{author}{\bibfnamefont{S.}~\bibnamefont{Stringari}},
  \bibinfo{journal}{Rev. Mod. Phys.} \textbf{\bibinfo{volume}{71}},
  \bibinfo{pages}{463} (\bibinfo{year}{1999}).

\bibitem[{\citenamefont{Ketterle et~al.}(1999)\citenamefont{Ketterle, Durfee,
  and Stamper-Kurn}}]{Ketterle1999a}
\bibinfo{author}{\bibfnamefont{W.}~\bibnamefont{Ketterle}},
  \bibinfo{author}{\bibfnamefont{D.~S.} \bibnamefont{Durfee}},
  \bibnamefont{and} \bibinfo{author}{\bibfnamefont{D.~M.}
  \bibnamefont{Stamper-Kurn}}, in \emph{\bibinfo{booktitle}{Proceedings of the
  International School of Physics - Enrico Fermi}}, edited by
  \bibinfo{editor}{\bibfnamefont{M.}~\bibnamefont{Inguscio}},
  \bibinfo{editor}{\bibfnamefont{S.}~\bibnamefont{Stringari}},
  \bibnamefont{and} \bibinfo{editor}{\bibfnamefont{C.~E.} \bibnamefont{Wieman}}
  (\bibinfo{publisher}{IOS Press}, \bibinfo{year}{1999}),
  p.~\bibinfo{pages}{67}.

\bibitem[{\citenamefont{Ketterle}(1999)}]{Ketterle1999b}
\bibinfo{author}{\bibfnamefont{W.}~\bibnamefont{Ketterle}},
  \bibinfo{journal}{Phys. Today} \textbf{\bibinfo{volume}{52}},
  \bibinfo{pages}{30} (\bibinfo{year}{1999}).

\bibitem[{\citenamefont{Regal et~al.}(2004)\citenamefont{Regal, Greiner, and
  Jin}}]{Regal2004b}
\bibinfo{author}{\bibfnamefont{C.~A.} \bibnamefont{Regal}},
  \bibinfo{author}{\bibfnamefont{M.}~\bibnamefont{Greiner}}, \bibnamefont{and}
  \bibinfo{author}{\bibfnamefont{D.~S.} \bibnamefont{Jin}},
  \bibinfo{journal}{Phys. Rev. Lett.} \textbf{\bibinfo{volume}{92}},
  \bibinfo{pages}{040403} (\bibinfo{year}{2004}).

\bibitem[{\citenamefont{Stoof and Houbiers}(1999)}]{Stoof1999b}
\bibinfo{author}{\bibfnamefont{H.~T.~C.} \bibnamefont{Stoof}} \bibnamefont{and}
  \bibinfo{author}{\bibfnamefont{M.}~\bibnamefont{Houbiers}}, in
  \emph{\bibinfo{booktitle}{Proceedings of the International School of Physics
  ``Enrico Fermi'' on Bose-Einstein condensation in Varenna 1998}}, edited by
  \bibinfo{editor}{\bibfnamefont{M.}~\bibnamefont{Inguscio}},
  \bibinfo{editor}{\bibfnamefont{S.}~\bibnamefont{Stringari}},
  \bibnamefont{and} \bibinfo{editor}{\bibfnamefont{C.~E.} \bibnamefont{Wieman}}
  (\bibinfo{publisher}{IOS Press, Amsterdam}, \bibinfo{year}{1999}), p.
  \bibinfo{pages}{175}.

\bibitem[{\citenamefont{Grimm}(2007)}]{Grimm2007a}
\bibinfo{author}{\bibfnamefont{R.}~\bibnamefont{Grimm}}, in
  \emph{\bibinfo{booktitle}{Proceedings of the International School of Physics
  - Enrico Fermi, Course CLXIV, Varenna, 2006}}, edited by
  \bibinfo{editor}{\bibfnamefont{M.}~\bibnamefont{Inguscio}},
  \bibinfo{editor}{\bibfnamefont{W.}~\bibnamefont{Ketterle}}, \bibnamefont{and}
  \bibinfo{editor}{\bibfnamefont{C.}~\bibnamefont{Salomon}}
  (\bibinfo{publisher}{eprint cond-mat/0703091}, \bibinfo{year}{2007}).

\bibitem[{\citenamefont{Gross}(1961)}]{Gross1961a}
\bibinfo{author}{\bibfnamefont{E.~P.} \bibnamefont{Gross}},
  \bibinfo{journal}{Nuovo Cim.} \textbf{\bibinfo{volume}{20}},
  \bibinfo{pages}{454} (\bibinfo{year}{1961}).

\bibitem[{\citenamefont{Pitaevskii}(1961)}]{Pitaevskii1961a}
\bibinfo{author}{\bibfnamefont{L.~P.} \bibnamefont{Pitaevskii}},
  \bibinfo{journal}{[Zh. Eksp. Teor. Fiz. 40, 646 (1961)] Sov. Phys. JETP}
  \textbf{\bibinfo{volume}{13}}, \bibinfo{pages}{451} (\bibinfo{year}{1961}).

\bibitem[{\citenamefont{Davis et~al.}(2002)\citenamefont{Davis, Morgan, and
  Burnett}}]{Davis2002b}
\bibinfo{author}{\bibfnamefont{M.~J.} \bibnamefont{Davis}},
  \bibinfo{author}{\bibfnamefont{S.~A.} \bibnamefont{Morgan}},
  \bibnamefont{and} \bibinfo{author}{\bibfnamefont{K.}~\bibnamefont{Burnett}},
  \bibinfo{journal}{Phys. Rev. A} \textbf{\bibinfo{volume}{66}},
  \bibinfo{pages}{053618} (\bibinfo{year}{2002}).

\bibitem[{\citenamefont{K{\"o}hl et~al.}(2002)\citenamefont{K{\"o}hl, Davis,
  Gardiner, H{\"a}nsch, and Esslinger}}]{Kohl2002a}
\bibinfo{author}{\bibfnamefont{M.}~\bibnamefont{K{\"o}hl}},
  \bibinfo{author}{\bibfnamefont{M.~J.} \bibnamefont{Davis}},
  \bibinfo{author}{\bibfnamefont{C.~W.} \bibnamefont{Gardiner}},
  \bibinfo{author}{\bibfnamefont{T.}~\bibnamefont{H{\"a}nsch}},
  \bibnamefont{and}
  \bibinfo{author}{\bibfnamefont{T.}~\bibnamefont{Esslinger}},
  \bibinfo{journal}{Phys. Rev. Lett.} \textbf{\bibinfo{volume}{88}},
  \bibinfo{pages}{080402} (\bibinfo{year}{2002}).

\bibitem[{\citenamefont{Davis and Gardiner}(2002)}]{Davis2002a}
\bibinfo{author}{\bibfnamefont{M.~J.} \bibnamefont{Davis}} \bibnamefont{and}
  \bibinfo{author}{\bibfnamefont{C.~W.} \bibnamefont{Gardiner}},
  \bibinfo{journal}{J. Phys. B} \textbf{\bibinfo{volume}{35}},
  \bibinfo{pages}{733} (\bibinfo{year}{2002}).

\bibitem[{\citenamefont{Stwalley}(1976)}]{Stwalley1976b}
\bibinfo{author}{\bibfnamefont{W.~C.} \bibnamefont{Stwalley}},
  \bibinfo{journal}{Phys. Rev. Lett.} \textbf{\bibinfo{volume}{37}},
  \bibinfo{pages}{1628} (\bibinfo{year}{1976}).

\bibitem[{\citenamefont{Tiesinga et~al.}(1992)\citenamefont{Tiesinga, Moerdijk,
  Verhaar, and Stoof}}]{Tiesinga1992a}
\bibinfo{author}{\bibfnamefont{E.}~\bibnamefont{Tiesinga}},
  \bibinfo{author}{\bibfnamefont{A.}~\bibnamefont{Moerdijk}},
  \bibinfo{author}{\bibfnamefont{B.~J.} \bibnamefont{Verhaar}},
  \bibnamefont{and} \bibinfo{author}{\bibfnamefont{H.~T.~C.}
  \bibnamefont{Stoof}}, \bibinfo{journal}{Phys. Rev. A}
  \textbf{\bibinfo{volume}{46}}, \bibinfo{pages}{R1167} (\bibinfo{year}{1992}).

\bibitem[{\citenamefont{Tiesinga et~al.}(1993)\citenamefont{Tiesinga, Verhaar,
  and Stoof}}]{Tiesinga1993a}
\bibinfo{author}{\bibfnamefont{E.}~\bibnamefont{Tiesinga}},
  \bibinfo{author}{\bibfnamefont{B.~J.} \bibnamefont{Verhaar}},
  \bibnamefont{and} \bibinfo{author}{\bibfnamefont{H.~T.~C.}
  \bibnamefont{Stoof}}, \bibinfo{journal}{Phys. Rev. A}
  \textbf{\bibinfo{volume}{47}}, \bibinfo{pages}{4114} (\bibinfo{year}{1993}).

\bibitem[{\citenamefont{Burnett}(1998)}]{Burnett1998a}
\bibinfo{author}{\bibfnamefont{K.}~\bibnamefont{Burnett}},
  \bibinfo{journal}{Nature (London)} \textbf{\bibinfo{volume}{392}},
  \bibinfo{pages}{125} (\bibinfo{year}{1998}).

\bibitem[{\citenamefont{K{\"o}hler et~al.}(2006)\citenamefont{K{\"o}hler,
  G{\'{o}}ral, and Julienne}}]{Kohler2006a}
\bibinfo{author}{\bibfnamefont{T.}~\bibnamefont{K{\"o}hler}},
  \bibinfo{author}{\bibfnamefont{K.}~\bibnamefont{G{\'{o}}ral}},
  \bibnamefont{and} \bibinfo{author}{\bibfnamefont{P.~S.}
  \bibnamefont{Julienne}}, \bibinfo{journal}{Rev. Mod. Phys.}
  \textbf{\bibinfo{volume}{78}}, \bibinfo{pages}{1311} (\bibinfo{year}{2006}).

\bibitem[{\citenamefont{Petrov et~al.}(2005)\citenamefont{Petrov, Salomon, and
  Shlyapnikov}}]{Petrov2005a}
\bibinfo{author}{\bibfnamefont{D.~S.} \bibnamefont{Petrov}},
  \bibinfo{author}{\bibfnamefont{C.}~\bibnamefont{Salomon}}, \bibnamefont{and}
  \bibinfo{author}{\bibfnamefont{G.~V.} \bibnamefont{Shlyapnikov}},
  \bibinfo{journal}{Phys. Rev. A} \textbf{\bibinfo{volume}{71}},
  \bibinfo{pages}{012708} (\bibinfo{year}{2005}).

\bibitem[{\citenamefont{Zwierlein et~al.}(2004)\citenamefont{Zwierlein, Stan,
  Schunck, Raupach, Kerman, and Ketterle}}]{Zwierlein2004a}
\bibinfo{author}{\bibfnamefont{M.~W.} \bibnamefont{Zwierlein}},
  \bibinfo{author}{\bibfnamefont{C.~A.} \bibnamefont{Stan}},
  \bibinfo{author}{\bibfnamefont{C.~H.} \bibnamefont{Schunck}},
  \bibinfo{author}{\bibfnamefont{S.~M.~F.} \bibnamefont{Raupach}},
  \bibinfo{author}{\bibfnamefont{A.~J.} \bibnamefont{Kerman}},
  \bibnamefont{and} \bibinfo{author}{\bibfnamefont{W.}~\bibnamefont{Ketterle}},
  \bibinfo{journal}{Phys. Rev. Lett.} \textbf{\bibinfo{volume}{92}},
  \bibinfo{pages}{120403} (\bibinfo{year}{2004}).

\bibitem[{\citenamefont{Bartenstein et~al.}(2004)\citenamefont{Bartenstein,
  Altmeyer, Riedl, Jochim, Chin, Denschlag, and Grimm}}]{Bartenstein2004a}
\bibinfo{author}{\bibfnamefont{M.}~\bibnamefont{Bartenstein}},
  \bibinfo{author}{\bibfnamefont{A.}~\bibnamefont{Altmeyer}},
  \bibinfo{author}{\bibfnamefont{S.}~\bibnamefont{Riedl}},
  \bibinfo{author}{\bibfnamefont{S.}~\bibnamefont{Jochim}},
  \bibinfo{author}{\bibfnamefont{C.}~\bibnamefont{Chin}},
  \bibinfo{author}{\bibfnamefont{J.~H.} \bibnamefont{Denschlag}},
  \bibnamefont{and} \bibinfo{author}{\bibfnamefont{R.}~\bibnamefont{Grimm}},
  \bibinfo{journal}{Phys. Rev. Lett.} \textbf{\bibinfo{volume}{92}},
  \bibinfo{pages}{120401} (\bibinfo{year}{2004}).

\bibitem[{\citenamefont{Schmiedmayer and Folman}(2001)}]{Schmiedmayer2000a}
\bibinfo{author}{\bibfnamefont{J.}~\bibnamefont{Schmiedmayer}}
  \bibnamefont{and} \bibinfo{author}{\bibfnamefont{R.}~\bibnamefont{Folman}},
  \bibinfo{journal}{C. R. Acad. Sci. Paris IV} \textbf{\bibinfo{volume}{2}},
  \bibinfo{pages}{333} (\bibinfo{year}{2001}).

\bibitem[{\citenamefont{Pitaevskii and Stringari}(2003)}]{Pitaevskii2003a}
\bibinfo{author}{\bibfnamefont{L.~P.} \bibnamefont{Pitaevskii}}
  \bibnamefont{and}
  \bibinfo{author}{\bibfnamefont{S.}~\bibnamefont{Stringari}},
  \emph{\bibinfo{title}{{B}ose-{E}instein Condensation}}
  (\bibinfo{publisher}{Clarendon Press}, \bibinfo{year}{2003}).

\bibitem[{\citenamefont{Jaksch et~al.}(1998)\citenamefont{Jaksch, Bruder,
  Cirac, Gardiner, and Zoller}}]{Jaksch1998a}
\bibinfo{author}{\bibfnamefont{D.}~\bibnamefont{Jaksch}},
  \bibinfo{author}{\bibfnamefont{C.}~\bibnamefont{Bruder}},
  \bibinfo{author}{\bibfnamefont{J.~I.} \bibnamefont{Cirac}},
  \bibinfo{author}{\bibfnamefont{C.~W.} \bibnamefont{Gardiner}},
  \bibnamefont{and} \bibinfo{author}{\bibfnamefont{P.}~\bibnamefont{Zoller}},
  \bibinfo{journal}{Phys. Rev. Lett.} \textbf{\bibinfo{volume}{81}},
  \bibinfo{pages}{3108} (\bibinfo{year}{1998}).

\bibitem[{\citenamefont{Bloch}(2004)}]{Bloch2004a}
\bibinfo{author}{\bibfnamefont{I.}~\bibnamefont{Bloch}},
  \bibinfo{journal}{Phys. World} \textbf{\bibinfo{volume}{17}},
  \bibinfo{pages}{25} (\bibinfo{year}{2004}).

\bibitem[{\citenamefont{van Oosten et~al.}(2001)\citenamefont{van Oosten,
  van~der Straten, and Stoof}}]{vanOosten2001a}
\bibinfo{author}{\bibfnamefont{D.}~\bibnamefont{van Oosten}},
  \bibinfo{author}{\bibfnamefont{P.}~\bibnamefont{van~der Straten}},
  \bibnamefont{and} \bibinfo{author}{\bibfnamefont{H.~T.~C.}
  \bibnamefont{Stoof}}, \bibinfo{journal}{Phys. Rev. A}
  \textbf{\bibinfo{volume}{63}}, \bibinfo{pages}{053601}
  (\bibinfo{year}{2001}).

\bibitem[{\citenamefont{Greiner et~al.}(2002)\citenamefont{Greiner, Mandel,
  Esslinger, H{\"a}nsch, and Bloch}}]{Greiner2002a}
\bibinfo{author}{\bibfnamefont{M.}~\bibnamefont{Greiner}},
  \bibinfo{author}{\bibfnamefont{O.}~\bibnamefont{Mandel}},
  \bibinfo{author}{\bibfnamefont{T.}~\bibnamefont{Esslinger}},
  \bibinfo{author}{\bibfnamefont{T.~W.} \bibnamefont{H{\"a}nsch}},
  \bibnamefont{and} \bibinfo{author}{\bibfnamefont{I.}~\bibnamefont{Bloch}},
  \bibinfo{journal}{Nature (London)} \textbf{\bibinfo{volume}{415}},
  \bibinfo{pages}{39} (\bibinfo{year}{2002}).

\bibitem[{\citenamefont{Tonks}(1936)}]{Tonks1936a}
\bibinfo{author}{\bibfnamefont{L.}~\bibnamefont{Tonks}},
  \bibinfo{journal}{Phys. Rev.} \textbf{\bibinfo{volume}{50}},
  \bibinfo{pages}{955} (\bibinfo{year}{1936}).

\bibitem[{\citenamefont{Girardeau}(1960)}]{Girardeau1960a}
\bibinfo{author}{\bibfnamefont{M.}~\bibnamefont{Girardeau}},
  \bibinfo{journal}{J. Math. Phys. (NY)} \textbf{\bibinfo{volume}{1}},
  \bibinfo{pages}{516} (\bibinfo{year}{1960}).

\bibitem[{\citenamefont{Paredes et~al.}(2004)\citenamefont{Paredes, Widera,
  Murg, Mandel, F{\"o}lling, Cirac, Shlyapnikov, H{\"a}nsch, and
  Bloch}}]{Paredes2004a}
\bibinfo{author}{\bibfnamefont{B.}~\bibnamefont{Paredes}},
  \bibinfo{author}{\bibfnamefont{A.}~\bibnamefont{Widera}},
  \bibinfo{author}{\bibfnamefont{V.}~\bibnamefont{Murg}},
  \bibinfo{author}{\bibfnamefont{O.}~\bibnamefont{Mandel}},
  \bibinfo{author}{\bibfnamefont{S.}~\bibnamefont{F{\"o}lling}},
  \bibinfo{author}{\bibfnamefont{I.}~\bibnamefont{Cirac}},
  \bibinfo{author}{\bibfnamefont{G.~V.} \bibnamefont{Shlyapnikov}},
  \bibinfo{author}{\bibfnamefont{T.~W.} \bibnamefont{H{\"a}nsch}},
  \bibnamefont{and} \bibinfo{author}{\bibfnamefont{I.}~\bibnamefont{Bloch}},
  \bibinfo{journal}{Nature (London)} \textbf{\bibinfo{volume}{429}},
  \bibinfo{pages}{277} (\bibinfo{year}{2004}).

\bibitem[{\citenamefont{Cornwall et~al.}(1974)\citenamefont{Cornwall, Jackiw,
  and Tomboulis}}]{Cornwall1974a}
\bibinfo{author}{\bibfnamefont{J.~M.} \bibnamefont{Cornwall}},
  \bibinfo{author}{\bibfnamefont{R.}~\bibnamefont{Jackiw}}, \bibnamefont{and}
  \bibinfo{author}{\bibfnamefont{E.}~\bibnamefont{Tomboulis}},
  \bibinfo{journal}{Phys. Rev. D} \textbf{\bibinfo{volume}{10}},
  \bibinfo{pages}{2428} (\bibinfo{year}{1974}).

\bibitem[{\citenamefont{Luttinger and Ward}(1960)}]{Luttinger1960a}
\bibinfo{author}{\bibfnamefont{J.~M.} \bibnamefont{Luttinger}}
  \bibnamefont{and} \bibinfo{author}{\bibfnamefont{J.~C.} \bibnamefont{Ward}},
  \bibinfo{journal}{Phys. Rev.} \textbf{\bibinfo{volume}{118}},
  \bibinfo{pages}{1417} (\bibinfo{year}{1960}).

\bibitem[{\citenamefont{Baym}(1962)}]{Baym1962a}
\bibinfo{author}{\bibfnamefont{G.}~\bibnamefont{Baym}}, \bibinfo{journal}{Phys.
  Rev.} \textbf{\bibinfo{volume}{127}}, \bibinfo{pages}{1391}
  (\bibinfo{year}{1962}).

\bibitem[{\citenamefont{Berges}(2002{\natexlab{a}})}]{Berges2002a}
\bibinfo{author}{\bibfnamefont{J.}~\bibnamefont{Berges}},
  \bibinfo{journal}{Nucl. Phys.} \textbf{\bibinfo{volume}{A699}},
  \bibinfo{pages}{847} (\bibinfo{year}{2002}{\natexlab{a}}).

\bibitem[{\citenamefont{Aarts et~al.}(2002)\citenamefont{Aarts, Ahrensmeier,
  Baier, Berges, and Serreau}}]{Aarts2002b}
\bibinfo{author}{\bibfnamefont{G.}~\bibnamefont{Aarts}},
  \bibinfo{author}{\bibfnamefont{D.}~\bibnamefont{Ahrensmeier}},
  \bibinfo{author}{\bibfnamefont{R.}~\bibnamefont{Baier}},
  \bibinfo{author}{\bibfnamefont{J.}~\bibnamefont{Berges}}, \bibnamefont{and}
  \bibinfo{author}{\bibfnamefont{J.}~\bibnamefont{Serreau}},
  \bibinfo{journal}{Phys. Rev. D} \textbf{\bibinfo{volume}{66}},
  \bibinfo{pages}{045008} (\bibinfo{year}{2002}).

\bibitem[{\citenamefont{Gasenzer et~al.}(2005)\citenamefont{Gasenzer, Berges,
  Schmidt, and Seco}}]{Gasenzer2005a}
\bibinfo{author}{\bibfnamefont{T.}~\bibnamefont{Gasenzer}},
  \bibinfo{author}{\bibfnamefont{J.}~\bibnamefont{Berges}},
  \bibinfo{author}{\bibfnamefont{M.~G.} \bibnamefont{Schmidt}},
  \bibnamefont{and} \bibinfo{author}{\bibfnamefont{M.}~\bibnamefont{Seco}},
  \bibinfo{journal}{Phys. Rev. A} \textbf{\bibinfo{volume}{72}},
  \bibinfo{pages}{063604} (\bibinfo{year}{2005}).

\bibitem[{\citenamefont{Rey et~al.}(2004)\citenamefont{Rey, Hu, Calzetta,
  Roura, and Clark}}]{Rey2004a}
\bibinfo{author}{\bibfnamefont{A.}~\bibnamefont{Rey}},
  \bibinfo{author}{\bibfnamefont{B.}~\bibnamefont{Hu}},
  \bibinfo{author}{\bibfnamefont{E.}~\bibnamefont{Calzetta}},
  \bibinfo{author}{\bibfnamefont{A.}~\bibnamefont{Roura}}, \bibnamefont{and}
  \bibinfo{author}{\bibfnamefont{C.}~\bibnamefont{Clark}},
  \bibinfo{journal}{Phys. Rev. A} \textbf{\bibinfo{volume}{69}},
  \bibinfo{pages}{033610} (\bibinfo{year}{2004}).

\bibitem[{\citenamefont{Temme and Gasenzer}(2006)}]{Temme2006a}
\bibinfo{author}{\bibfnamefont{K.}~\bibnamefont{Temme}} \bibnamefont{and}
  \bibinfo{author}{\bibfnamefont{T.}~\bibnamefont{Gasenzer}},
  \bibinfo{journal}{Phys. Rev. A} \textbf{\bibinfo{volume}{74}},
  \bibinfo{pages}{053603} (\bibinfo{year}{2006}).

\bibitem[{\citenamefont{Berges and Serreau}(2003)}]{Berges2003b}
\bibinfo{author}{\bibfnamefont{J.}~\bibnamefont{Berges}} \bibnamefont{and}
  \bibinfo{author}{\bibfnamefont{J.}~\bibnamefont{Serreau}},
  \bibinfo{journal}{Phys. Rev. Lett.} \textbf{\bibinfo{volume}{91}},
  \bibinfo{pages}{111601} (\bibinfo{year}{2003}).

\bibitem[{\citenamefont{Cooper et~al.}(2003)\citenamefont{Cooper, Dawson, and
  Mihaila}}]{Cooper2003a}
\bibinfo{author}{\bibfnamefont{F.}~\bibnamefont{Cooper}},
  \bibinfo{author}{\bibfnamefont{J.~F.} \bibnamefont{Dawson}},
  \bibnamefont{and} \bibinfo{author}{\bibfnamefont{B.}~\bibnamefont{Mihaila}},
  \bibinfo{journal}{Phys. Rev. D} \textbf{\bibinfo{volume}{67}},
  \bibinfo{pages}{056003} (\bibinfo{year}{2003}).

\bibitem[{\citenamefont{Arrizabalaga et~al.}(2004)\citenamefont{Arrizabalaga,
  Smit, and Tranberg}}]{Arrizabalaga2004a}
\bibinfo{author}{\bibfnamefont{A.}~\bibnamefont{Arrizabalaga}},
  \bibinfo{author}{\bibfnamefont{J.}~\bibnamefont{Smit}}, \bibnamefont{and}
  \bibinfo{author}{\bibfnamefont{A.}~\bibnamefont{Tranberg}},
  \bibinfo{journal}{JHEP} \textbf{\bibinfo{volume}{10}}, \bibinfo{pages}{017}
  (\bibinfo{year}{2004}).

\bibitem[{\citenamefont{Berges et~al.}(2003)\citenamefont{Berges,
  Bors{\'{a}}nyi, and Serreau}}]{Berges2003a}
\bibinfo{author}{\bibfnamefont{J.}~\bibnamefont{Berges}},
  \bibinfo{author}{\bibfnamefont{S.}~\bibnamefont{Bors{\'{a}}nyi}},
  \bibnamefont{and} \bibinfo{author}{\bibfnamefont{J.}~\bibnamefont{Serreau}},
  \bibinfo{journal}{Nucl. Phys.} \textbf{\bibinfo{volume}{B660}},
  \bibinfo{pages}{51} (\bibinfo{year}{2003}).

\bibitem[{\citenamefont{Berges et~al.}(2004)\citenamefont{Berges,
  Bors{\'{a}}nyi, and Wetterich}}]{Berges2004b}
\bibinfo{author}{\bibfnamefont{J.}~\bibnamefont{Berges}},
  \bibinfo{author}{\bibfnamefont{S.}~\bibnamefont{Bors{\'{a}}nyi}},
  \bibnamefont{and}
  \bibinfo{author}{\bibfnamefont{C.}~\bibnamefont{Wetterich}},
  \bibinfo{journal}{Phys. Rev. Lett.} \textbf{\bibinfo{volume}{93}},
  \bibinfo{pages}{142002} (\bibinfo{year}{2004}).

\bibitem[{\citenamefont{Berges}(2005)}]{Berges2005a}
\bibinfo{author}{\bibfnamefont{J.}~\bibnamefont{Berges}},
  \bibinfo{journal}{eprint hep-ph/0409233; AIP Conf. Proc.}
  \textbf{\bibinfo{volume}{739}}, \bibinfo{pages}{3} (\bibinfo{year}{2005}).

\bibitem[{\citenamefont{Aarts and Berges}(2002)}]{Aarts2002a}
\bibinfo{author}{\bibfnamefont{G.}~\bibnamefont{Aarts}} \bibnamefont{and}
  \bibinfo{author}{\bibfnamefont{J.}~\bibnamefont{Berges}},
  \bibinfo{journal}{Phys. Rev. Lett.} \textbf{\bibinfo{volume}{88}},
  \bibinfo{pages}{041603} (\bibinfo{year}{2002}).

\bibitem[{\citenamefont{Berges}(2002{\natexlab{b}})}]{Berges:2002wf}
\bibinfo{author}{\bibfnamefont{J.}~\bibnamefont{Berges}},
  \bibinfo{journal}{Nucl. Phys.} \textbf{\bibinfo{volume}{A702}},
  \bibinfo{pages}{351} (\bibinfo{year}{2002}{\natexlab{b}}),
  \eprint{hep-ph/0201204}.

\bibitem[{\citenamefont{Hopf}(1952)}]{Hopf1952a}
\bibinfo{author}{\bibfnamefont{E.}~\bibnamefont{Hopf}}, \bibinfo{journal}{J.
  Ratl. Mech. Anal.} \textbf{\bibinfo{volume}{1}}, \bibinfo{pages}{87}
  (\bibinfo{year}{1952}).

\bibitem[{\citenamefont{Martin et~al.}(1973)\citenamefont{Martin, Siggia, and
  Rose}}]{Martin1973a}
\bibinfo{author}{\bibfnamefont{P.~C.} \bibnamefont{Martin}},
  \bibinfo{author}{\bibfnamefont{E.~D.} \bibnamefont{Siggia}},
  \bibnamefont{and} \bibinfo{author}{\bibfnamefont{H.~A.} \bibnamefont{Rose}},
  \bibinfo{journal}{Phys. Rev. A} \textbf{\bibinfo{volume}{8}},
  \bibinfo{pages}{423} (\bibinfo{year}{1973}).

\bibitem[{\citenamefont{Hohenberg and Halperin}(1977)}]{Hohenberg1977a}
\bibinfo{author}{\bibfnamefont{P.~C.} \bibnamefont{Hohenberg}}
  \bibnamefont{and} \bibinfo{author}{\bibfnamefont{B.~I.}
  \bibnamefont{Halperin}}, \bibinfo{journal}{Rev. Mod. Phys.}
  \textbf{\bibinfo{volume}{49}}, \bibinfo{pages}{435} (\bibinfo{year}{1977}).

\bibitem[{\citenamefont{Phythian}(1975)}]{Phythian1975a}
\bibinfo{author}{\bibfnamefont{R.}~\bibnamefont{Phythian}},
  \bibinfo{journal}{J. Phys. A} \textbf{\bibinfo{volume}{8}},
  \bibinfo{pages}{1423} (\bibinfo{year}{1975}).

\bibitem[{\citenamefont{{De D}ominicis}(1976)}]{DeDominicis1976a}
\bibinfo{author}{\bibfnamefont{C.}~\bibnamefont{{De D}ominicis}},
  \bibinfo{journal}{J. Phys. (Paris) C} \textbf{\bibinfo{volume}{1}},
  \bibinfo{pages}{247} (\bibinfo{year}{1976}).

\bibitem[{\citenamefont{Janssen}(1976)}]{Janssen1976a}
\bibinfo{author}{\bibfnamefont{H.-K.} \bibnamefont{Janssen}},
  \bibinfo{journal}{Z. Phys. B} \textbf{\bibinfo{volume}{23}},
  \bibinfo{pages}{377} (\bibinfo{year}{1976}).

\bibitem[{\citenamefont{Bausch et~al.}(1976)\citenamefont{Bausch, Janssen, and
  Wagner}}]{Bausch1976a}
\bibinfo{author}{\bibfnamefont{R.}~\bibnamefont{Bausch}},
  \bibinfo{author}{\bibfnamefont{H.~K.} \bibnamefont{Janssen}},
  \bibnamefont{and} \bibinfo{author}{\bibfnamefont{H.}~\bibnamefont{Wagner}},
  \bibinfo{journal}{Z. Phys. B} \textbf{\bibinfo{volume}{24}},
  \bibinfo{pages}{113} (\bibinfo{year}{1976}).

\bibitem[{\citenamefont{Phythian}(1977)}]{Phythian1977a}
\bibinfo{author}{\bibfnamefont{R.}~\bibnamefont{Phythian}},
  \bibinfo{journal}{J. Phys. A} \textbf{\bibinfo{volume}{10}},
  \bibinfo{pages}{777} (\bibinfo{year}{1977}).

\bibitem[{\citenamefont{{De D}ominicis and Peliti}(1978)}]{DeDominicis1978a}
\bibinfo{author}{\bibfnamefont{C.}~\bibnamefont{{De D}ominicis}}
  \bibnamefont{and} \bibinfo{author}{\bibfnamefont{L.}~\bibnamefont{Peliti}},
  \bibinfo{journal}{Phys. Rev. B} \textbf{\bibinfo{volume}{18}},
  \bibinfo{pages}{353} (\bibinfo{year}{1978}).

\bibitem[{\citenamefont{Schwinger}(1961)}]{Schwinger1961a}
\bibinfo{author}{\bibfnamefont{J.}~\bibnamefont{Schwinger}},
  \bibinfo{journal}{J. Math. Phys.} \textbf{\bibinfo{volume}{2}},
  \bibinfo{pages}{407} (\bibinfo{year}{1961}).

\bibitem[{\citenamefont{Keldysh}(1964)}]{Keldysh1964a}
\bibinfo{author}{\bibfnamefont{L.~V.} \bibnamefont{Keldysh}},
  \bibinfo{journal}{[Sov. Phys. JETP {\bf 20}, 1018 (1965)] Zh. Eksp. Teor.
  Fiz.} \textbf{\bibinfo{volume}{47}}, \bibinfo{pages}{1515}
  (\bibinfo{year}{1964}).

\bibitem[{\citenamefont{Zhou et~al.}(1980)\citenamefont{Zhou, Su, Hao, and
  Yu}}]{Chou1980a}
\bibinfo{author}{\bibfnamefont{G.}~\bibnamefont{Zhou}},
  \bibinfo{author}{\bibfnamefont{Z.}~\bibnamefont{Su}},
  \bibinfo{author}{\bibfnamefont{B.}~\bibnamefont{Hao}}, \bibnamefont{and}
  \bibinfo{author}{\bibfnamefont{L.}~\bibnamefont{Yu}}, \bibinfo{journal}{Phys.
  Rev. B} \textbf{\bibinfo{volume}{22}}, \bibinfo{pages}{3385}
  (\bibinfo{year}{1980}).

\bibitem[{\citenamefont{Chou et~al.}(1985)\citenamefont{Chou, Su, Hao, and
  Yu}}]{Chou1985a}
\bibinfo{author}{\bibfnamefont{G.}~\bibnamefont{Chou}},
  \bibinfo{author}{\bibfnamefont{Z.}~\bibnamefont{Su}},
  \bibinfo{author}{\bibfnamefont{B.}~\bibnamefont{Hao}}, \bibnamefont{and}
  \bibinfo{author}{\bibfnamefont{L.}~\bibnamefont{Yu}}, \bibinfo{journal}{Phys.
  Rep.} \textbf{\bibinfo{volume}{118}}, \bibinfo{pages}{1}
  (\bibinfo{year}{1985}).

\bibitem[{\citenamefont{Blagoev et~al.}(2001)\citenamefont{Blagoev, Cooper,
  Dawson, and Mihaila}}]{Blagoev2001a}
\bibinfo{author}{\bibfnamefont{K.~B.} \bibnamefont{Blagoev}},
  \bibinfo{author}{\bibfnamefont{F.}~\bibnamefont{Cooper}},
  \bibinfo{author}{\bibfnamefont{J.~F.} \bibnamefont{Dawson}},
  \bibnamefont{and} \bibinfo{author}{\bibfnamefont{B.}~\bibnamefont{Mihaila}},
  \bibinfo{journal}{Phys. Rev. D} \textbf{\bibinfo{volume}{64}},
  \bibinfo{pages}{125003} (\bibinfo{year}{2001}).

\bibitem[{\citenamefont{Aarts and Smit}(1997)}]{Aarts1997a}
\bibinfo{author}{\bibfnamefont{G.}~\bibnamefont{Aarts}} \bibnamefont{and}
  \bibinfo{author}{\bibfnamefont{J.}~\bibnamefont{Smit}},
  \bibinfo{journal}{Phys. Lett. B} \textbf{\bibinfo{volume}{393}},
  \bibinfo{pages}{395} (\bibinfo{year}{1997}).

\bibitem[{\citenamefont{Buchm{\"u}ller and
  Jakov{\'{a}}c}(1997)}]{Buchmuller1997a}
\bibinfo{author}{\bibfnamefont{W.}~\bibnamefont{Buchm{\"u}ller}}
  \bibnamefont{and}
  \bibinfo{author}{\bibfnamefont{A.}~\bibnamefont{Jakov{\'{a}}c}},
  \bibinfo{journal}{Phys. Lett. B} \textbf{\bibinfo{volume}{407}},
  \bibinfo{pages}{39} (\bibinfo{year}{1997}).

\bibitem[{\citenamefont{Aarts and Smit}(1998)}]{Aarts1998a}
\bibinfo{author}{\bibfnamefont{G.}~\bibnamefont{Aarts}} \bibnamefont{and}
  \bibinfo{author}{\bibfnamefont{J.}~\bibnamefont{Smit}},
  \bibinfo{journal}{Nucl. Phys. B} \textbf{\bibinfo{volume}{511}},
  \bibinfo{pages}{451} (\bibinfo{year}{1998}).

\bibitem[{\citenamefont{Cooper et~al.}(2001)\citenamefont{Cooper, Khare, and
  Rose}}]{Cooper2001a}
\bibinfo{author}{\bibfnamefont{F.}~\bibnamefont{Cooper}},
  \bibinfo{author}{\bibfnamefont{A.}~\bibnamefont{Khare}}, \bibnamefont{and}
  \bibinfo{author}{\bibfnamefont{H.}~\bibnamefont{Rose}},
  \bibinfo{journal}{Phys. Lett. B} \textbf{\bibinfo{volume}{515}},
  \bibinfo{pages}{463} (\bibinfo{year}{2001}).

\bibitem[{\citenamefont{Wetterich}(1997)}]{Wetterich:1997rp}
\bibinfo{author}{\bibfnamefont{C.}~\bibnamefont{Wetterich}},
  \bibinfo{journal}{Phys. Rev. E} \textbf{\bibinfo{volume}{56}},
  \bibinfo{pages}{2687} (\bibinfo{year}{1997}), \eprint{hep-th/9703006}.

\bibitem[{\citenamefont{Jeon}(2005)}]{Jeon2005a}
\bibinfo{author}{\bibfnamefont{S.}~\bibnamefont{Jeon}}, \bibinfo{journal}{Phys.
  Rev. C} \textbf{\bibinfo{volume}{72}}, \bibinfo{pages}{014907}
  (\bibinfo{year}{2005}).

\bibitem[{\citenamefont{Hartree}(1928)}]{Hartree1928a}
\bibinfo{author}{\bibfnamefont{D.~R.} \bibnamefont{Hartree}},
  \bibinfo{journal}{Proc. Cambridge Phil. Soc.} \textbf{\bibinfo{volume}{24}},
  \bibinfo{pages}{89} (\bibinfo{year}{1928}).

\bibitem[{\citenamefont{Fock}(1930)}]{Fock1930a}
\bibinfo{author}{\bibfnamefont{V.}~\bibnamefont{Fock}}, \bibinfo{journal}{Z.
  Phys.} \textbf{\bibinfo{volume}{61}}, \bibinfo{pages}{126}
  (\bibinfo{year}{1930}).

\bibitem[{\citenamefont{Bogoliubov}(1947)}]{Bogoliubov1947a}
\bibinfo{author}{\bibfnamefont{N.~N.} \bibnamefont{Bogoliubov}},
  \bibinfo{journal}{J. Phys. (USSR)} \textbf{\bibinfo{volume}{11}},
  \bibinfo{pages}{23} (\bibinfo{year}{1947}).

\bibitem[{\citenamefont{Aarts and Berges}(2001)}]{Aarts2001a}
\bibinfo{author}{\bibfnamefont{G.}~\bibnamefont{Aarts}} \bibnamefont{and}
  \bibinfo{author}{\bibfnamefont{J.}~\bibnamefont{Berges}},
  \bibinfo{journal}{Phys. Rev. D} \textbf{\bibinfo{volume}{64}},
  \bibinfo{pages}{105010} (\bibinfo{year}{2001}).

\bibitem[{\citenamefont{Fetter and Walecka}(1971)}]{Fetter1971a}
\bibinfo{author}{\bibfnamefont{A.~L.} \bibnamefont{Fetter}} \bibnamefont{and}
  \bibinfo{author}{\bibfnamefont{J.~D.} \bibnamefont{Walecka}},
  \emph{\bibinfo{title}{Quantum theory of many-particle systems}}
  (\bibinfo{publisher}{MacGraw-Hill, New York}, \bibinfo{year}{1971}).

\bibitem[{\citenamefont{Berges and Cox}(2001)}]{Berges2001a}
\bibinfo{author}{\bibfnamefont{J.}~\bibnamefont{Berges}} \bibnamefont{and}
  \bibinfo{author}{\bibfnamefont{J.}~\bibnamefont{Cox}},
  \bibinfo{journal}{Phys. Lett.} \textbf{\bibinfo{volume}{B517}},
  \bibinfo{pages}{369} (\bibinfo{year}{2001}).

\bibitem[{\citenamefont{Berges}(2004)}]{Berges2004a}
\bibinfo{author}{\bibfnamefont{J.}~\bibnamefont{Berges}},
  \bibinfo{journal}{Phys. Rev. D} \textbf{\bibinfo{volume}{70}},
  \bibinfo{pages}{105010} (\bibinfo{year}{2004}).

\end{thebibliography}

\end{document}